\documentclass[structabstract]{aa} 
\usepackage{graphicx}
\usepackage{array}
\usepackage{amssymb}
\usepackage{natbib}
\usepackage{float}
\usepackage{color}
\bibpunct{(}{)}{;}{a}{,}{,}

\newcommand{\degree}{\ensuremath{^\circ}}
\begin{document}
   \title{Magnetic field structure around cores with very low luminosity objects}

   \subtitle{}

   \author{Soam, A.\inst{1}
          \and
          Maheswar, G.\inst{1}
          \and
          Chang Won Lee\inst{2,7}
          \and
          Sami Dib\inst{3,4}
          \and
		  Bhatt, H. C.\inst{5}
		  \and
		  Motohide Tamura \inst{6}
		  \and Gwanjeong Kim \inst{2,7}
          }

   \institute{Aryabhatta Research Institute of Observational Sciences (ARIES), Nainital 263002, India.
              \email{archana@aries.res.in}
         	 \and
             Korea Astronomy $\&$ Space Science Institute (KASI), 776 Daedeokdae-ro, Yuseong-gu, Daejeon, Republic of Korea.
             \and
             Niels Bohr International Academy, Niels Bohr Institute, Blegdamsvej 17, DK-2100, Copenhagen, Denmark 
             \and 
             Centre for Star and Planet Formation, University of Copenhagen, {\O}ster Voldgade 5-7., DK-1350, Copenhagen, Denmark           
	         \and 
	         Indian Institute of Astrophysics, Kormangala (IIA), Bangalore 560034, India.   
	         \and 
	         National Astronomical Observatory of Japan (NAOJ), Mitaka, Tokyo 181-8588, Japan
	         \and
	         University of Science \& Technology, 217 Gajungro, Yuseong-gu, Daejeon, 305-333, Korea
             }

   \date{Received......, Accepted.....}

% \abstract{}{}{}{}{} 
% 5 {} token are mandatory
 
  \abstract
   {}
   { We carried out optical polarimetry of five dense cores, (IRAM 04191, L1521F, L328, L673-7, and L1014) which are found to harbour very low luminosity objects (VeLLOs; L$_{int}\lesssim 0.1 L_{\odot}$). This study was conducted mainly to understand the role played by the magnetic field in the formation of very low and substellar mass range objects.}
   {Light from the stars, while passing through the dust grains that are aligned with their short axis parallel to an external magnetic field, becomes linearly polarised. The polarisation position angles measured for the stars can provide the plane-of-the sky magnetic field orientation. Because the light in the optical wavelength range is most efficiently polarised by the dust grains typically found at the outer layers of the molecular clouds, optical polarimetry mostly traces the magnetic field orientation of the core envelope.}
   {The polarisation observations of stars projected on IRAM 04191, L328, L673-7, and L1014 were obtained in the R-band and those of L1521F were obtained in the V-band. The angular offsets between the envelope magnetic field direction (inferred from optical polarisation measurements) and the outflow position angles from the VeLLOs in IRAM 04191, L1521F, L328, L673-7, and L1014 are found to be 84$\degree$, 53$\degree$, 24$\degree$, 08$\degree$, and 15$\degree$, respectively. The mean value of the offsets for all the five clouds is $\sim37\degree$. If we exclude IRAM 04191, the mean value reduces to become $\sim25\degree$. In IRAM 04191, the offset between the projected envelope and the inner magnetic field (inferred from the submillimetre data obtained using SCUBA-POL) is found to be $\sim68\degree$. The inner magnetic field, however, is found to be nearly aligned with the projected position angles of the minor axis, the rotation axis of the cloud, and the outflow from the IRAM 04191-IRS. We discuss a possible explanation for the nearly perpendicular orientation between the envelope and core scale magnetic fields in IRAM04191. The angular offset between the envelope magnetic field direction and the minor axis of IRAM 04191, L1521F, L673-7, and L1014 are 82$\degree$, 60$\degree$, 47$\degree$, and 55$\degree$, respectively. The mean value of the offsets between the envelope magnetic field and the minor axis position angles for the four cores is found to be $\sim60\degree$.}
   {The results obtained from our study on the limited sample of five cores with VeLLOs show that the outflows in three of them tend to nearly align with the envelope magnetic field.}


   \maketitle
%
%________________________________________________________________

%#############################             INTRODUCTION

\section{Introduction}

The influence of magnetic fields in all stages of cloud and star formation is still unclear. In a magnetic field dominated scenario for isolated low-mass star formation, the cores are envisaged to gradually condense out of a magnetically subcritical background cloud, through ambipolar diffusion \citep[e.g. ][]{1987ARA&A..25...23S, 1993prpl.conf..327M, 1999osps.conf..305M, 2003ApJ...599..363A}. The material, mediated by the magnetic field lines, settles into a disk-like morphology of a few thousand AU in size. This would result in the cloud scale magnetic field lines becoming parallel to the symmetry axis of the flattened region. The magnetic field inside the infall radius is expected to show a pinched hourglass morphology \citep{1993ApJ...415..680F, 1993ApJ...417..243G}. The magnetic field lines just outside this region are expected to be moderately pinched, and to gradually join to the smooth and uniform field distribution expected at the subcritical background envelope. On the contrary, in scenarios where magnetic fields play a lesser role, cores are thought to form at the intersection of supersonic turbulent flows \citep{2004RvMP...76..125M,2007ApJ...661..262D, 2010ApJ...723..425D}. Consequently, the distribution of the magnetic field lines is expected to be more chaotic \citep{2004Ap&SS.292..225C}. One can put these models to test by mapping magnetic field morphology at different physical scales of the clouds. The relationship between the field morphology and other properties, such as the structure, kinematics, and alignment of any bipolar outflows that may be present in the cloud could be examined. 

The magnetic field maps of the molecular clouds are made using polarisation measurements of background starlight in optical wavelengths. The observed polarisation is caused by the selective extinction of the light as it passes through elongated dust grains that are aligned with their minor axis parallel to the local magnetic fields of the cloud. The exact mechanisms by which the alignment of the dust grains with the magnetic field occurs are still unclear \citep{2003JQSRT..79..881L, 2004ASPC..309..467R}. The polarisation position angles in optical wavelengths trace the plane-of-the-sky orientation of the ambient magnetic field at the periphery of the molecular clouds \citep[with $A_{V}$ $\sim$ 1-2 mag; ][]{1995ApJ...448..748G, 1996ASPC...97..325G}. Polarisation observations in near-IR and submillimetre/millimetre wavelengths are used to trace the field lines in more dense regions of the cloud \citep[e.g. ][]{2000ApJ...537L.135W, 2011ASPC..449..207T} that are opaque to optical polarisation measurements. The line-of-sight component of the magnetic field is measured using Zeeman observations \citep[e.g. ][]{1993ApJ...407..175C}.

In this work, we present the results of optical polarisation measurements of stars projected on the fields containing five dense cores namely, IRAM 04191+1522 (IRAM 04191, hereafter), L1521F, L328, L673-7, and L1014\footnote{L1014 is the core L1014-2 which contains the VeLLO, L1014-IRS. This is a core with a $2^{\prime}$ angular extent. Another cloud, L1014-1, is located $\sim$ $10^{\prime}$ north-east of L1014-2. The cloud L1014-1 is also known as B362.}. Very low luminosity objects (VeLLOs; L$_{int}\lesssim 0.1 L_{\odot}$) are detected in these cores based on the data from the \textit{Spitzer Space Telescope} (SST). The VeLLOs are interesting sources as their luminosity is an order of magnitude lower than the accretion luminosity $L_{acc}$ $\sim$ 1.6 $L_{\odot}$ expected for a 0.08$M_{\odot}$ protostar with an accretion rate of $\sim$ $10^{-6}$ $M_{\odot}$ $yr^{-1}$ \citep{1987IAUS..115..417S} and 3$R_{\odot}$ stellar radius. It has been speculated that these sources are either progenitors of brown dwarfs \citep[e.g. ][]{2013ApJ...777...50L} or very low mass protostars. Detailed studies of these VeLLOs have been carried out to understand their properties (properties relevant to this study are summarised in Table \ref{tab:parameters}). These VeLLOs are classified as class 0 objects/candidate proto-brown dwarfs, and are also found to show outflow activities. The outflows from the protostars are thought to influence their surrounding environment by generating turbulence which could scramble a relatively weak magnetic field in their vicinity. This could weaken any initial alignment between the core magnetic field and the envelope magnetic field. The estimated outflow parameters suggest that VeLLOs have the most compact, lowest mass, and the least energetic outflows compared to known Class 0/I outflows from low-mass stars \citep[e.g. ][]{2002A&A...393..927B, 2004A&A...426..503W, 2006ApJ...649L..37B, 2011ApJ...743..201P}. We therefore expect the outflows from these sources to have the least significant effect on their surroundings. This would enable the regions to preserve the initial condition that may have existed prior to the initiation of star formation. By studying the magnetic field structure around the dense cores with VeLLOs, we expect to understand the role played by the magnetic field in the formation of very low-mass stars or substellar objects. This paper is structured in the following manner. The \S.~\ref{sec:obs,datared} describes the procedure adopted for the data acquisition, reduction and the determination of the degree of polarisation and the polarisation position angles. The polarisation results of the background stars for all the five clouds are presented in \S.~\ref{RESULTS}. The analysis and discussion of the results obtained are given in \S.~\ref{Discussion}. Our conclusions are summarised in \S.~\ref{Conclusions}.
  
%==============
\begin{table*}
\caption{\bf Basic properties of the sources studied. References are given in parentheses.}\label{tab:parameters}
\begin{tabular}{cccccc}
\hline\hline
\multicolumn{6}{c}{{\bf Cloud properties}}\\
Information				&IRAM 04191		&L1521F			&L328			&L673-7			&L1014			\\
\hline
 Right Ascension$^{\dagger\dagger}$	(J2000)		& 04 21 57	&04 28 39.8		&18 16 59.5  	&19 20 23		&21 24 07.5		\\  
 Declination$^{\dagger\dagger}$		(J2000)		& $+$15 29 46	&$+$26 51 35	&$-$18 02 03 	&$+$11 22 49.5	&$+$49 59 05 	\\	
Distance (pc)$^{\dagger}$	&127$\pm$25		&136$\pm$36		&217$\pm$30		&240$\pm$45		&258$\pm$50		\\
Galactic plane (Deg)        &139            &135            &29             &28             &57             \\
Minor axis P.A. ($\theta_{min}$ in Deg)	&30 (1)	&82 (2)		&--			&0	(2)			&70	(2)			\\
\bf Velocity dispersion ($\Delta v$, km$s^{-1}$) & $1.50\pm0.04$ (3) & $1.26\pm0.02$ (3) & $1.30\pm0.02$ (3) & $0.88\pm0.04$ (3)  & $0.93\pm0.01$ (3) \\
\hline
\multicolumn{6}{c}{\bf VeLLO properties}\\
Outflow P.A. ($\theta_{out}$ in Deg)			&28 (1)			&75 (4)			&20 (5)			&55 (6)			&30 (7)	\\
Source classification      &class 0 (8) 	&class 0 (4)   &proto-brown dwarf (5)     &class 0 (9)   & class 0 (10)\\
\hline
\end{tabular}

{\bf References:} 1. \citet{2002A&A...393..927B}, 2. \citet{2008A&A...487..993K}, 3. \citet[in preparation;][]{inpreparation}, 4. \citet{2013ApJ...774...20T}, 5. \citet{2013ApJ...777...50L}, 6. \citet{2010ApJ...721..995D}, 7. \citet{2005ApJ...633L.129B}, 8. \citet{2006ApJ...651..945D}, 9. \citet{2008ApJS..179..249D}, 10. \citet{2004ApJS..154..396Y}, $^{\dagger}$ \citet{2010A&A...509A..44M}, $^{\dagger\dagger}$ The core centers are the positions of the peak emission in CO obtained from \citet{2005ApJ...619..948L} (IRAM 04191), \citet{2004A&A...420..957C} (L1521F), \citet{2009ApJ...693.1290L} (L328), \citet{2002AJ....124.2756V} (L673-7), and \citet{2005IAUS..235P.193C} (L1014).
\end{table*}      %cloud and VeLLO properties
%==============

%##############################                OBSERVATIONS AND DATA REDUCTION
\section{Observations and data reduction}\label{sec:obs,datared}
Polarimetric observations of stars projected on five dense cores were carried out using the Aries IMaging POLarimeter \citep[AIMPOL,][]{2004BASI...32..159R} mounted at the cassegrain focus of the 104 cm Sampurnanand telescope of Aryabhatta Research Institute of Observational Sciences (ARIES), Nainital, India, coupled with TK 1024 pixels$\times$1024 pixels CCD camera. Of these, only the central 325 pixels$\times$325 pixels are used for the observations. AIMPOL consists of an achromatic half-wave plate (HWP) modulator and a Wollaston prism beam-splitter. The observations of the stars projected on three dense cores namely IRAM 04191, L673-7 and L1014 were carried out in standard Johnson $R_{c}$ filter having $\lambda_{R_{eff}}$=0.630$\mu$m. Observations of L328 were carried out in standard Johnson $R_{kc}$=0.760$\mu$m filter. The observations of stars towards L1521F were carried out in standard V filter of 0.550$\mu$m. The mean exposure time per observed field was $\sim$ 200s. The plate scale of the CCD used is 1.48$^{\prime\prime}/$pixel and the field of view is $\sim 8^{\prime}$ in diameter. The full width at half maximum (FWHM) varies from 2 to 3 pixels. The read out noise and gain of CCD are 7.0 $e^{-}$  and 11.98 $e^{-}$/ADU. Table \ref{tab:obslog} gives the log of the observations.

The dual-beam polarizing prism allows us to measure the polarisation by simultaneously imaging both orthogonal polarisation states onto the detector. Both the half-wave plate fast axis and the axis of the Wollaston prism are kept normal to the optical axis of the system. The intensities of the two beams, extraordinary and ordinary, are given by
\begin{equation}
I_{e}(\alpha)=\frac{I_{unpol}}{2}+I_{pol}\times cos^{2}(\theta-2\alpha)
\end{equation}
\begin{equation}
I_{o}(\alpha)=\frac{I_{unpol}}{2}+I_{pol}\times sin^{2}(\theta-2\alpha) ,
\end{equation} 
where $I_{pol}$ and $I_{unpol}$ are the polarised and unpolarised intensities. The position angles $\theta$ and $\alpha$ are the angles for which polarisation vector and the half-wave plate fast axis are measured with respect to the axis of the Wollaston prism. The values {\it $I_{o}$} and {\it $I_{e}$} are the fluxes of ordinary and extraordinary beams. 

The standard aperture photometry in the IRAF package was used to extract the fluxes of ordinary ({\it $I_{o}$}) and extraordinary ({\it $I_{e}$}) beams for all the observed sources with a good signal-to-noise ratio. The ratio {\it {R($\alpha$)}} is given by
 \begin{equation}
 R(\alpha) = \frac{\frac{{I_{e}}(\alpha)}{{I_{o}}(\alpha)}-1} {\frac{I_{e}(\alpha)} {I_{o}(\alpha)}+1} =  P\times cos(2\theta - 4\alpha) ,
\end{equation} 
where {\it P} is the fraction of total linearly polarised light and $\theta$ is the polarisation angle of the plane of polarisation. The {\it $\alpha$} is the position of the fast axis of the HWP at $0\degree$, $22.5\degree$, $45\degree$, and $67.5\degree$ corresponding to four normalised Stokes parameters, respectively, q[R($0\degree$)], u[R($22.5\degree$)], $q_{1}$[R($45\degree$)], and $u_{1}$[R($67.5\degree$)]. The errors in the normalised Stokes parameters ($\sigma_R$)($\alpha$)($\sigma_q$, $\sigma_u$, $\sigma_{q1}$, $\sigma_{u1}$) in per cent were estimated using the relation \citep{1998A&AS..128..369R}
\begin{equation}
\sigma_R(\alpha)= \frac{\sqrt{N_{e}+N_{o}+2N_{b}}}{N_{e}+N_{o}} ,
\end{equation} 
where $N_{o}$ and $N_{e}$ are the counts in ordinary and extraordinary beams, and $N_{b}$[= ({$N_{be}$}$+${$N_{bo}$})/2] is the average background counts around the extraordinary and ordinary rays. 

%==============
\begin{table*}
  \caption{Log of the optical polarimetric observations.}\label{tab:obslog}
  \begin{tabular}{lll}\hline
%  \begin{tabular}{p{1.3cm}p{5cm}p{1cm}}\hline
  Cloud ID          &  Date of observations & Filter\\
                    &  (year, month, date)  &        \\
\hline
 IRAM 04191         & 2010 Nov 14; 2011 Jan 26;                 &  $R_{c}$ \\
                    & 2011 Feb 27	                            &           \\
 L1521F             & 2007 Nov 01$-$2007 Nov 04       		    &  V     \\
 L328               & 2013 May 15, 2013 May 16                  &  $R_{kc}$  \\
 L673-7             & 2012 May 16, 2012 May 19        			&  $R_{c}$ \\                   
 L1014              & 2010 Nov 14; 2011 Nov 22            	    &  $R_{c}$ \\                                    
\hline
\end{tabular}
\end{table*}		%Observation Log
%==============

Zero polarisation standard stars were observed to check for any possible instrumental polarisation. The typical instrumental polarisation is found to be less than $\sim0.1\%$. The instrumental polarisation of AIMPOL on the 104 cm Sampurnanand Telescope has been monitored since 2004 for various observing programmes and found to be stable \citep[see ][]{2004BASI...32..159R, 2011MNRAS.411.1418E, 2013MNRAS.432.1502S}. The reference direction of the polariser was determined by observing polarised standard stars from \citet{1992AJ....104.1563S}. The results are presented in Table \ref{tab:std}.
We observed these un-polarised and polarised standards using the standard Johnson $R_{c}$ and $R_{kc}$ filters. \citet{1992AJ....104.1563S} used Kron-Cousins R filter for the observations of the standard stars. We found a good correlation of the observed values with the standard values (see Table \ref{tab:std}). The zero point offset was corrected using the offset seen between the standard position angle values given in \citet{1992AJ....104.1563S} and those obtained by us.

%==============
\begin{table}
\caption{Results of observed polarised standard stars.}\label{tab:std}
\begin{tabular}{lll}\hline
Date of     &P $\pm$ $\sigma_P$ 	&  $\theta$ $\pm$ $\sigma_{\theta}$  \\
Obs.		&(\%)            		& ($\degree$)                           \\\hline
\multicolumn{3}{l}{{\bf HD 236633}}\\
\multicolumn{3}{l}{$^\dagger$Standard values: 5.38 $\pm$ 0.02\%, 93.04 $\pm$ 0.15$\degree$}\\
01 Nov 2007 & 5.5 $\pm$ 0.1     & 92 $\pm$ 1\\
04 Nov 2007 & 5.6 $\pm$ 0.1     & 93 $\pm$ 1 \\
22 Nov 2011	& 4.8 $\pm$ 0.2     & 92 $\pm$ 1 \\
26 Nov 2011	& 5.3 $\pm$ 0.1     & 92 $\pm$ 1 \\
20 Dec 2011	& 5.3 $\pm$ 0.2     & 93 $\pm$ 2 \\
24 Dec 2011	& 5.7 $\pm$ 0.2     & 93 $\pm$ 1 \\\hline
\multicolumn{3}{l}{{\bf HD 236954}}\\
\multicolumn{3}{l}{$^\ddagger$Standard values: 6.16 $\pm$ 0.17\%, 110.0 $\pm$ 0.8$\degree$}\\
20 Dec 2011 & 6.0 $\pm$ 0.1     & 111 $\pm$ 2\\
20 Dec 2011 & 6.2 $\pm$ 0.4   	& 111 $\pm$ 2\\\hline
\multicolumn{3}{l}{{\bf BD$+$59$\degree$389}} \\
\multicolumn{3}{l}{$^\dagger$Standard values: 6.43 $\pm$ 0.02\%, 98.14 $\pm$ 0.10$\degree$}\\
03 Nov 2007	& 6.6 $\pm$ 0.1     & 98 $\pm$ 1 \\
04 Nov 2007 & 6.9 $\pm$ 0.1     & 98 $\pm$ 1 \\
14 Nov 2010 & 6.1 $\pm$ 0.1     & 97 $\pm$ 1 \\
26 Nov 2011 & 7.0 $\pm$ 0.2     & 98 $\pm$ 1  \\
20 Dec 2011 & 6.2 $\pm$ 0.1     & 98 $\pm$ 1  \\
24 Dec 2011 & 6.4 $\pm$ 0.1     & 98 $\pm$ 1 \\\hline
\multicolumn{3}{l}{{\bf HD 25443 }}\\
\multicolumn{3}{l}{$^\dagger$Standard values: 4.73 $\pm$ 0.05\%, 133.65$\pm$ 0.28$\degree$}\\
14 Nov 2010 & 4.9 $\pm$ 0.1     & 133 $\pm$ 2 \\
27 Feb 2011 & 4.9 $\pm$ 0.1     & 133 $\pm$ 1  \\ \hline
\multicolumn{3}{l}{{\bf BD$+$64$\degree$3106}}\\
\multicolumn{3}{l}{$^\dagger$Standard values: 5.69 $\pm$ 0.04\%, 96.63 $\pm$ 0.18$\degree$}\\
01 Nov 2007 & 5.6 $\pm$ 0.1     & 96 $\pm$ 1  \\
04 Nov 2007 & 5.3 $\pm$ 0.1     & 96 $\pm$ 1   \\\hline
\multicolumn{3}{l}{{\bf HD 19820}}\\
\multicolumn{3}{l}{$^\dagger$Standard values: 4.53 $\pm$ 0.02\%, 114.46$\pm$ 0.16$\degree$}\\
26 Jan 2011 & 4.5 $\pm$ 0.1     & 114 $\pm$ 1  \\\hline
\multicolumn{3}{l}{{\bf HD 245310}}\\
\multicolumn{3}{l}{$^\dagger$Standard values: 4.55 $\pm$ 0.06\%, 145.97$\pm$ 0.40$\degree$}\\
27 Feb 2011 & 4.4 $\pm$ 0.1     & 145 $\pm$ 1  \\\hline
\multicolumn{3}{l}{{\bf HD 154445}}\\
\multicolumn{3}{l}{$^\dagger$Standard values: 3.68 $\pm$ 0.07\%, 88.91 $\pm$ 0.58$\degree$}\\
16 May 2012 & 3.7 $\pm$ 0.1     & 98 $\pm$ 1  \\
19 May 2012 & 3.7 $\pm$ 0.1     & 98 $\pm$ 1  \\
20 May 2012 & 3.8 $\pm$ 0.1     & 99 $\pm$ 1 \\\hline
\multicolumn{3}{l}{{\bf HD 155197}}\\
\multicolumn{3}{l}{$^\dagger$Standard values: 4.27$\pm$ 0.03\%, 102.88 $\pm$ 0.18$\degree$}\\
16 May 2012 & 4.3 $\pm$ 0.1     & 110 $\pm$ 1  \\
19 May 2012 & 4.3 $\pm$ 0.1     & 111 $\pm$ 1 \\
20 May 2012 & 4.3 $\pm$ 0.1     & 111 $\pm$ 1  \\
{$^\ast$15 May 2013} & 4.2 $\pm$ 0.2     & 109 $\pm$ 1  \\\hline
\multicolumn{3}{l}{{\bf HD 161056}}\\
\multicolumn{3}{l}{$^\dagger$Standard values: 4.01$\pm$ 0.03\%, 67.33 $\pm$ 0.23$\degree$}\\
$^\ast$15 May 2013 & 3.8 $\pm$ 0.1     & 75 $\pm$ 1  \\\hline
\end{tabular}

$^\dagger$ In $R_{c}$ band from \citet{1992AJ....104.1563S} \\
$\ddagger$ R band values calculated using the Serkowski's law and V band results obtained from the \citet[][Original reference \citet{1956ApJS....2..389H}]{2000AJ....119..923H}\\
$^\ast$ Data for target stars and standards stars on 15 May 2013 was taken in Johnson $R_{kc}$ band \citep{1992AJ....104.1563S} \\
 
\end{table}		% Standard star observations
%==============

\section{Results}\label{RESULTS}

Results of our polarisation measurements of the stars projected on the fields of IRAM 04191, L1521F, L328, L6737, and L1014 are given in Table \ref{tab:pol_res}. Columns 1, 2, 3, 4, and 5 give the star identification, the Right Ascension (J2000), the declination (J2000), degree of polarisation (P) in per cent, and the polarisation position angles ($\theta_{p}$) in degree, respectively. The values of $\theta_{p}$ are measured towards the east from the north. We have chosen only those measurements for which the ratio of the P and the error in P ($\sigma_{P}$), P/$\sigma_{P}$, is $\geq$ 2. Additional information on the stars observed by us obtained from the literature are presented in the notes to Table \ref{tab:pol_res}.

Observationally, the polarisation measurements of stars provide the plane-of-the-sky magnetic field direction averaged over the entire line of sight weighted by the density and the alignment efficiency of the dust grains, but in order to obtain the orientation of the magnetic field of a cloud, in principle, the contribution due to the foreground dust component from the measured polarisation values has to be removed. However, from our previous studies on clouds located at distances less than $\sim$500 pc \citep[e.g. ][]{2011MNRAS.411.1418E, 2013MNRAS.432.1502S} and from the literature \citep{2009ApJ...704..891L}, the change in the values of the $\theta_{p}$ after correcting for the foreground contribution is not very significant especially when the observed P is relatively high ($\gtrsim 1\%$). The five clouds studied here are all located at distances $\lesssim250$ pc (Table \ref{tab:parameters}). Therefore in this work, we have not removed the foreground dust contribution from our polarisation measurements. 

Inside a molecular cloud, many of the molecule forming reactions occur on the surface of dust grains which could result in  completely different grain characteristics in terms of their shape, size and/or composition. Many studies have shown that the dust grains located inside the denser regions of molecular clouds are bigger than those located at the periphery of the clouds \citep{1980ApJ...235..905W, 2003AJ....126.1888K, 2005ASPC..343..321W, 2010A&A...522A..84O}. As dust grains with sizes similar to those found in the outer parts of molecular clouds are most efficient at polarizing light in the optical wavelengths, the maps of magnetic field geometry made using optical polarimetry used in this work are of the low density parts (envelope) of the dense cores \citep{1996ASPC...97..325G}.

\subsection{IRAM 04191}

%########################################################################################
\begin{figure}
\centering
\resizebox{8.5cm}{8.5cm}{\includegraphics{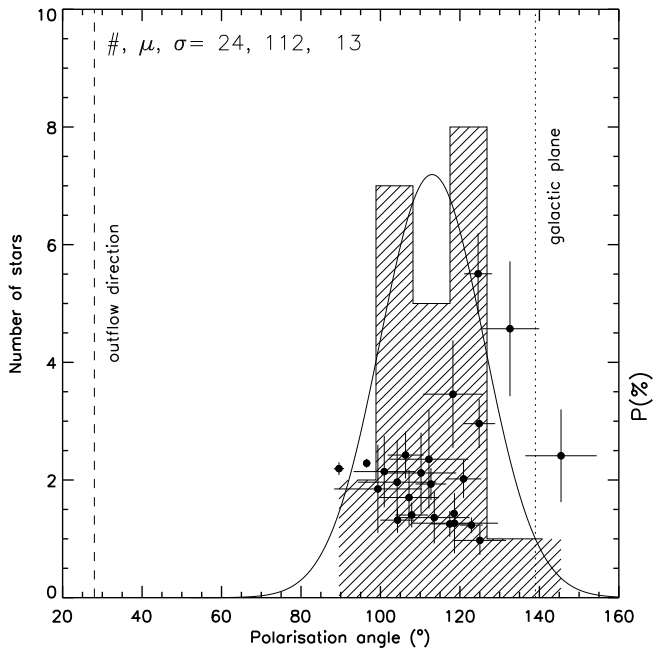}}
\caption{P versus $\theta_{p}$ of stars projected on IRAM 04191 excluding star 31 (Young Stellar Object (YSO)). The histogram of the $\theta_{p}$ with binsize 10$\degree$ is also presented. The solid curve represents a fit to the histogram of $\theta_{p}$. The position angle of the CO outflow associated with IRAM 04191-IRS is shown as a broken line. The position angle of the Galactic plane at the latitude of the cloud is shown as a dotted line.}\label{Fig:PvsPAiram}
\end{figure}
%########################################################################################
\begin{figure}
\resizebox{8.5cm}{8.5cm}{\includegraphics{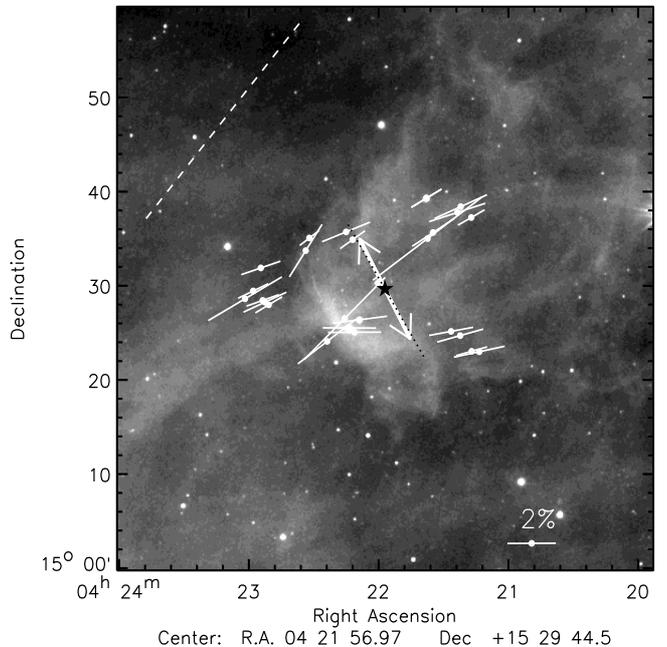}}
\caption{Polarisation vectors of the stars observed towards IRAM 04191 are shown on $1^\circ$ $\times$ $1^\circ$ WISE 12$\mu$m images of the region containing the cloud. The position of IRAM 04191-IRS is identified by a black star. The outflow (solid line with double arrow head) and the Galactic position angle (broken line) are also drawn. Mean value of submm polarisation angles is shown with a black dotted line. A vector with 2\% polarisation is shown as a reference.}\label{Fig:dss_wiseIRAM}
\end{figure}
%########################################################################################
 %=====================================================
 \begin{figure}
\resizebox{8cm}{8cm}{\includegraphics{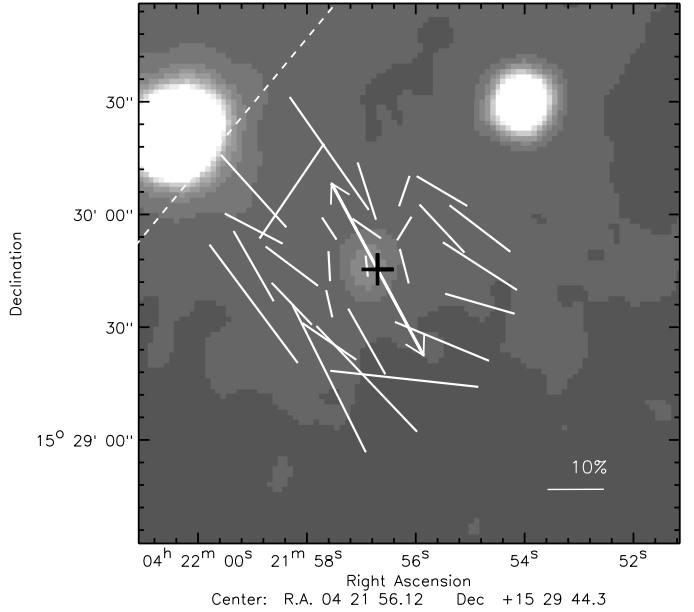}}
\caption{Magnetic field directions in IRAM 04191 interpreted based on submillimetre polarisation vectors obtained using SCUPOL \citep{2009ApJS..182..143M} are over-plotted on the WISE 4.6$\mu$m image of IRAM 04191. The polarisation position angles are rotated by 90$\degree$ to represent the magnetic field directions. The position of IRAM 04191-IRS is identified by a plus symbol. The Galactic plane is shown with a dashed line. A vector with 10\% polarisation is shown as a reference.}\label{Fig:scuIRAM}
\end{figure}
%=====================================================
\begin{figure}
\resizebox{8.5cm}{8.5cm}{\includegraphics{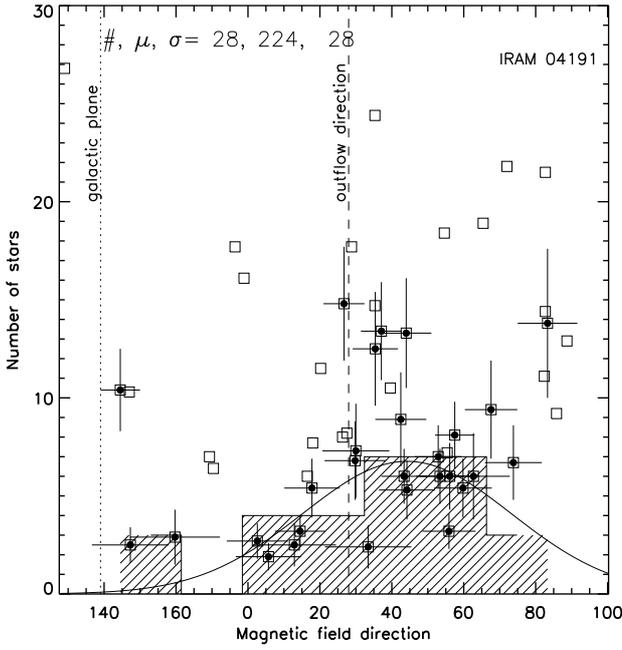}}
\caption{The plot shows the histogram of magnetic field position angles with binsize 10$\degree$ obtained from SCUPOL (a 90$\degree$ offset has been made to the SCUPOL polarisation position angles to obtain the magnetic field orientation). The plot also shows the distribution of P versus the magnetic field position angles. The open squares represent the data from the entire core which includes the regions containing both IRAM 04191-IRS and IRAS 04191+1523. The filled circles enclosed in open squares represent the measurements from a circular region of radius of 40$^{\prime\prime}$ about IRAM 04191-IRS. The position angle of the CO outflow associated with IRAM 04191-IRS is shown by a dashed line. The direction of the Galactic plane at the latitude of this cloud is shown by a dotted line.}\label{Fig:iram_scuba}
\end{figure}
%=====================================================
\begin{figure}
\resizebox{8.5cm}{8.5cm}{\includegraphics{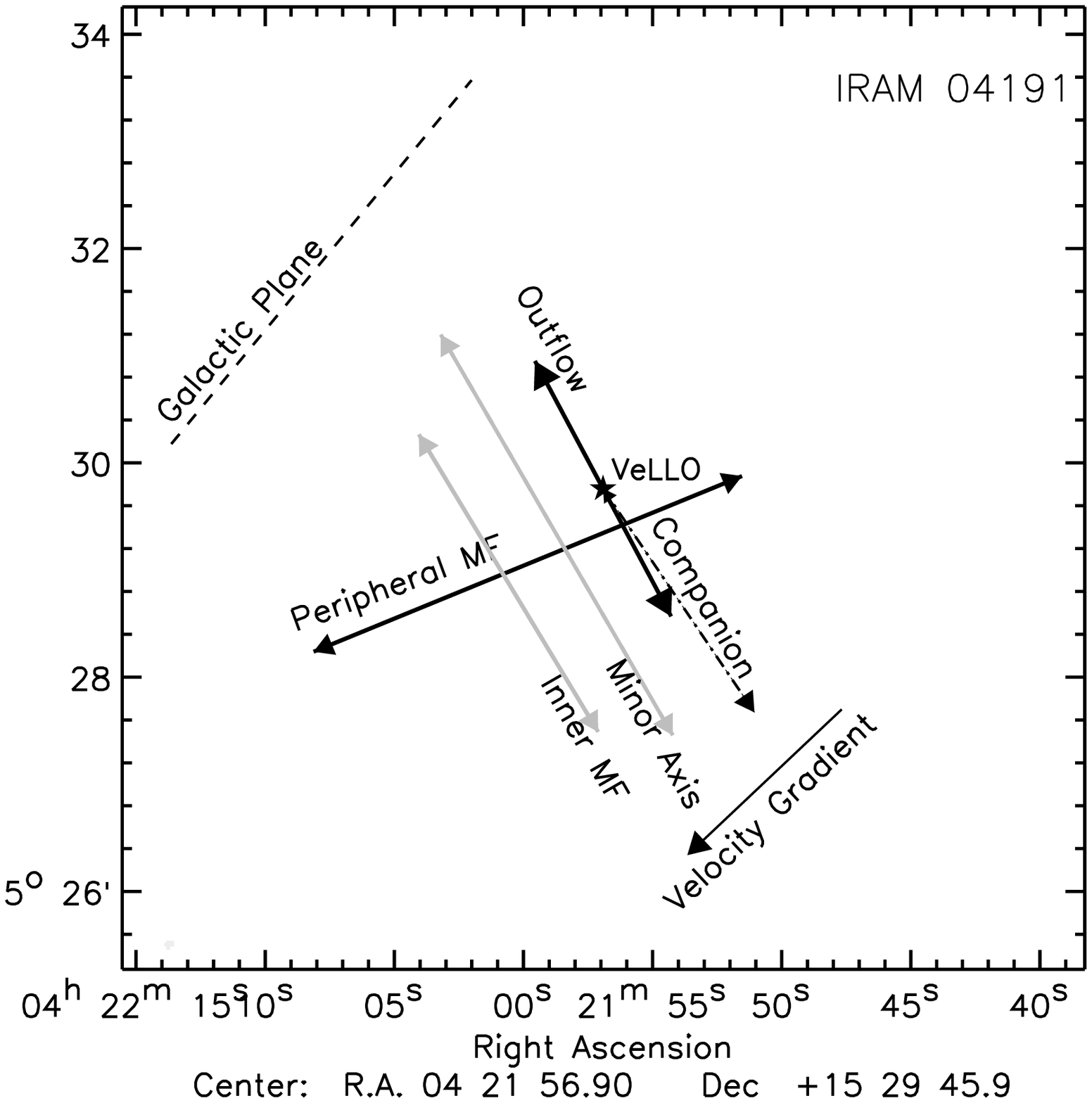}}
\caption{Schematic diagram of IRAM 04191 illustrating the orientations of the mean peripheral magnetic field (inferred from optical polarisation), mean direction of inner magnetic field (inferred by submillimetre polarisation), minor axis, position of IRAM 04191-IRS (star symbol), orientation of the companion source, direction of outflow from IRAM 04191-IRS, direction of velocity gradient, and the Galactic plane with respect to the north.}\label{Fig:schem}
\end{figure}
%=====================================================
A class 0 source, IRAM 04191-IRS was first detected in IRAM 04191 by \citet{2006ApJ...651..945D} at wavelengths shortward of 60$\mu$m using the data from the SST, although it was recognised as a class 0 source much earlier \citep{1999ApJ...513L..57A}. The P versus $\theta_{p}$ plot for the 25 stars that we observed are shown in Fig. \ref{Fig:PvsPAiram}. In Fig. \ref{Fig:dss_wiseIRAM}, we show the polarisation vectors of the stars we observed overlaid on the WISE 12$\mu$m image ($1^\circ$ $\times$ $1^\circ$) of the cloud. The mean values of P and $\theta_{p}$ and the corresponding standard deviations obtained from a Gaussian fit to the data are found to be 2.6$\pm$1.0\% and 112$\pm$13$\degree$. Among the sources that we observed, source 14 is identified as a YSO by \cite{2011ApJS..196....4R}. The YSOs show intrinsic polarisation due to scattering of radiation from the stars and the circumstellar dust envelopes \citep{1982A&AS...48..153B}. The observed values of P and $\theta_{p}$ for this source are 2.2$\pm$0.1\% and 90$\pm$1$\degree$. The sources that could possibly have intrinsic polarisation are need to be removed while estimating the local magnetic field properties. The mean values of P and $\theta_{p}$ after removing the source 14 are found to be 2.1$\pm$1.0\% and 112$\pm$13$\degree$.

The submillimetre polarisation measurements for IRAM 04191, made using SCUBA-POL\footnote{Submillimetre Common-User Bolometer Array polarimeter on James Clark Maxwell Telescope.} (now onwards SCUPOL), are available in the catalogue made by \citet{2009ApJS..182..143M}. The polarisation vectors are sampled on a 10$^{\prime\prime}$ grid. We have selected only those measurements for which the P/$\sigma_{P}$ $\geq$ 2. In Fig. \ref{Fig:scuIRAM}, the submillimetre polarisation vectors are plotted on the WISE 4.6$\mu$m image of IRAM 04191. The thermal continuum emission from aligned, elongated dust grains, measured in submillimetre wavelengths, is polarised and aligned to the longer axis of the grains. Therefore the submillimetre polarisation vectors have to be rotated by 90$\degree$ to infer the magnetic field orientation \citep{1996ASPC...97..325G, 2003ApJ...592..233W}. The P versus the magnetic field position angles (obtained after rotating $\theta_{p}$ values by 90$\degree$) are presented in Fig. \ref{Fig:iram_scuba}. The mean values of the P and the magnetic field position angles with corresponding standard deviations are found to be 13$\pm$7\% and 32$\pm$36$\degree$.

Towards the north-east of IRAM 04191-IRS and within an angular distance of $\sim1^{\prime}$ there is a class I source, IRAS 04191+1523. The submillimetre data provided by \citet{2009ApJS..182..143M} also include the region containing this source. In Fig. \ref{Fig:iram_scuba}, we show the measurements selected from a circular region of radius 40$^{\prime\prime}$ about IRAM 04191-IRS and include the measurements around IRAM 04191-IRS alone. The histogram of the magnetic field position angles is also presented in the same plot. The mean value of P for the measurements around IRAM 04191-IRS alone is found to be 7$\pm$4\%. The mean value of the magnetic field position angles around IRAM 04191-IRS alone obtained from a Gaussian fit is found to be 44$\pm$31$\degree$. This value represents the magnetic field orientation towards the denser, inner parts of the cloud. 

The presence of a highly collimated CO bipolar outflow was known to be associated with IRAM 04191-IRS \citep{1999ApJ...513L..57A}. From the CO map of the region presented by \citet{2002A&A...393..927B}, the estimated outflow direction is $28\degree$ with respect to the north (see Fig. \ref{Fig:PvsPAiram}). Orientations of both the polarisation vectors and the outflow are shown in Fig. \ref{Fig:dss_wiseIRAM}. Both IRAM 04191-IRS and IRAS 04191+1523 are found to have companion sources. The companion separation and orientation for IRAS 04191+1523 are found to be $\sim6^{\prime\prime}$ and $\sim304\degree$ \citep{2004A&A...427..651D}. The companion source for IRAM 04191-IRS was detected in the  high spatial resolution 1.3 mm continuum images obtained by \citet{2012ApJ...747L..43C}. They estimated an angular separation of $\sim8^{\prime\prime}$. Using the images provided by them, we determined a position angle of $\sim290\degree$ to the companion source. \citet{2012ApJ...747L..43C} also made a velocity map of the IRAM 04191, using the combined IRAM 30 m and PdBI N$_2$H$^+$ (1$-$0) data from \citet{2004A&A...419L..35B} and showed that there is a continuous velocity gradient across the dense core, increasing from north-west to south-east with a position angle of 133$\degree$. The centrosymmetric centroid-velocity curve along the direction of the mean velocity gradient is taken as strong evidence of the rotation of the core \citep{2013arXiv1305.0627B}. Thus, the rotation axis is oriented in a direction $\sim43\degree$ to the north. 

%{\bf Velocity gradient associated with the core rotation is seen in IRAM0 4191 shown in the maps produced by a dense gas tracer N$_2$H$^+$ \citep{2012ApJ...747L..43C}}. 

The orientation of the major axis of IRAM 04191 estimated by \citet{2002A&A...393..927B} based on the N$_{2}$H$^{+}$(1-0) intensity map of the IRAM 04191 protostellar envelope is found to be $120\degree$. Hence the minor axis position angle becomes $30\degree$. The Galactic plane at the latitude of the cloud is found to be $139\degree$ which is shown as a dashed line in both Fig. \ref{Fig:dss_wiseIRAM} and Fig. \ref{Fig:scuIRAM}. In Fig. \ref{Fig:schem}, we present a schematic diagram showing various orientations (see caption of Fig. \ref{Fig:schem} for detail).

%mean peripheral magnetic field (inferred from optical polarisation), mean direction of inner magnetic field (inferred by submillimetre polarisation), minor axis, position of IRAM 04191-IRS, orientation of the companion source (IRAM 04191+1523), outflow from IRAM 04191-IRS, direction of velocity gradient, and the Galactic plane with respect to the north. 

\subsection{L1521F}

%########################################################################################
\begin{figure}
\resizebox{8.5cm}{8.5cm}{\includegraphics{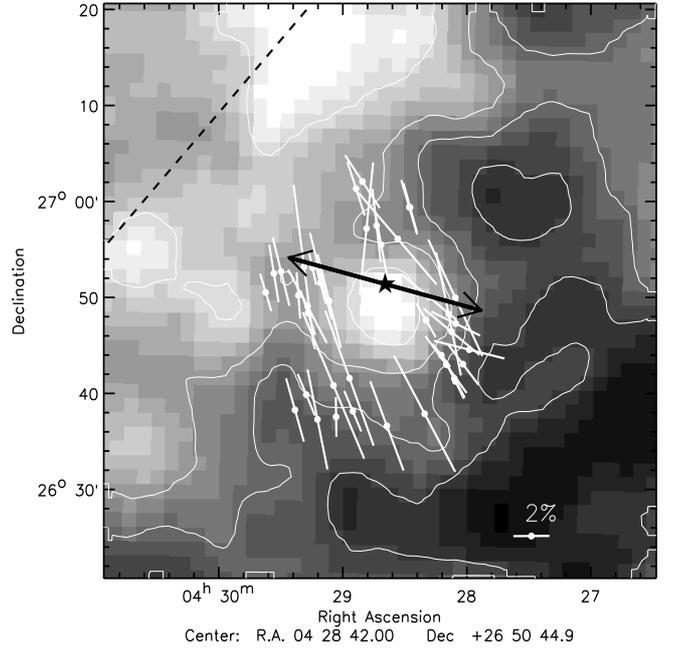}}
\caption{Polarisation vectors over-plotted on $1^\circ$ $\times$ $1^\circ$ IRAS 100$\mu$m image of the region containing the cloud L1521F. The position of L1521F-IRS is identified by a star. The outflow direction is shown as a thick line with double arrow heads. The broken line shows the Galactic plane orientation at the latitude of the cloud. The vector with 2$\%$ polarisation is shown for reference.}\label{Fig:dss_iras1521f}
\end{figure}
%#########################################################################################
\begin{figure}
\resizebox{8.5cm}{8.5cm}{\includegraphics{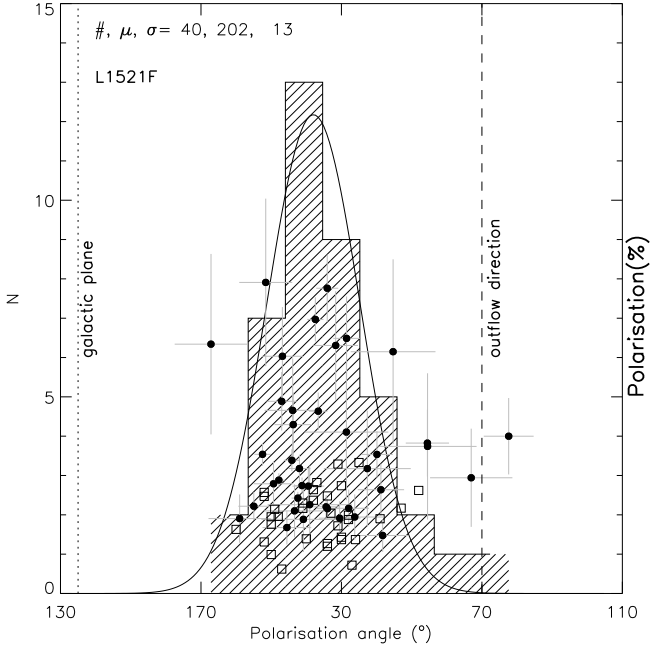}}
\caption{P versus the $\theta_{p}$ of stars projected on L1521F. The histogram of the $\theta_{p}$ with binsize 10$\degree$ is also shown. The position angle of the outflow direction is shown as a broken line. The direction of the Galactic plane at the latitude of this cloud is shown as a dotted line. The polarisation results of the stars within a circular region of 1$\degree$ around L1521F obtained from the \citet{2000AJ....119..923H} catalogue are shown as open square symbols. Filled circles represent the sources observed by us.}\label{Fig:PvsPAl1521f}
\end{figure}
%#########################################################################################
\begin{figure}
\resizebox{8.5cm}{8.5cm}{\includegraphics{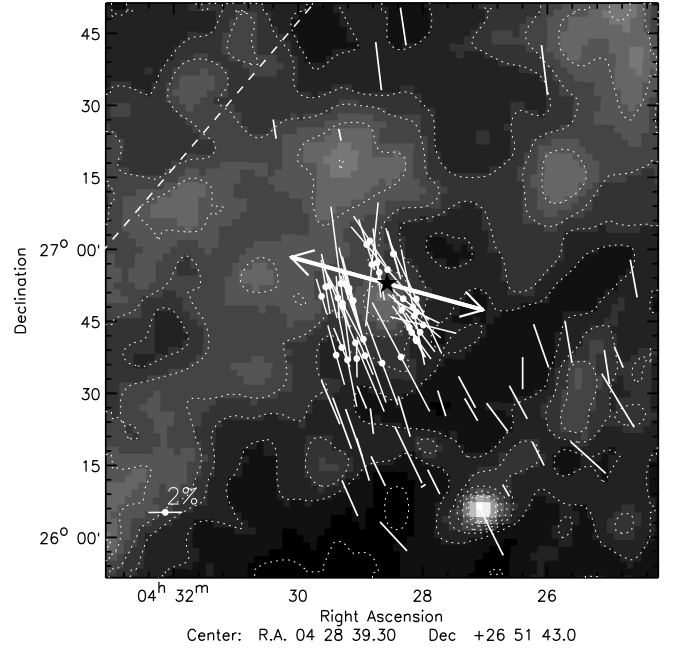}}
\caption{Optical polarisation vectors (lines with filled white circles) are over-plotted on a $2^\circ$ $\times$ $2^\circ$ IRAS 100$\mu$m image of L1521F. The polarisation vectors corresponding to the stars obtained from the \citet{2000AJ....119..923H} catalogue selected from a circular region of radius 1$\degree$ around L1521F are also plotted (lines without circles). Position of L1521F-IRS is shown with a star. The outflow direction is shown as a thick line with two headed arrow. The broken line shows the Galactic plane orientation at the latitude of the cloud. The vector with 2$\%$ polarisation is shown for reference.}\label{Fig:heilesL1521f}
\end{figure}
%#########################################################################################
\begin{figure}
\resizebox{8.0cm}{10cm}{\includegraphics{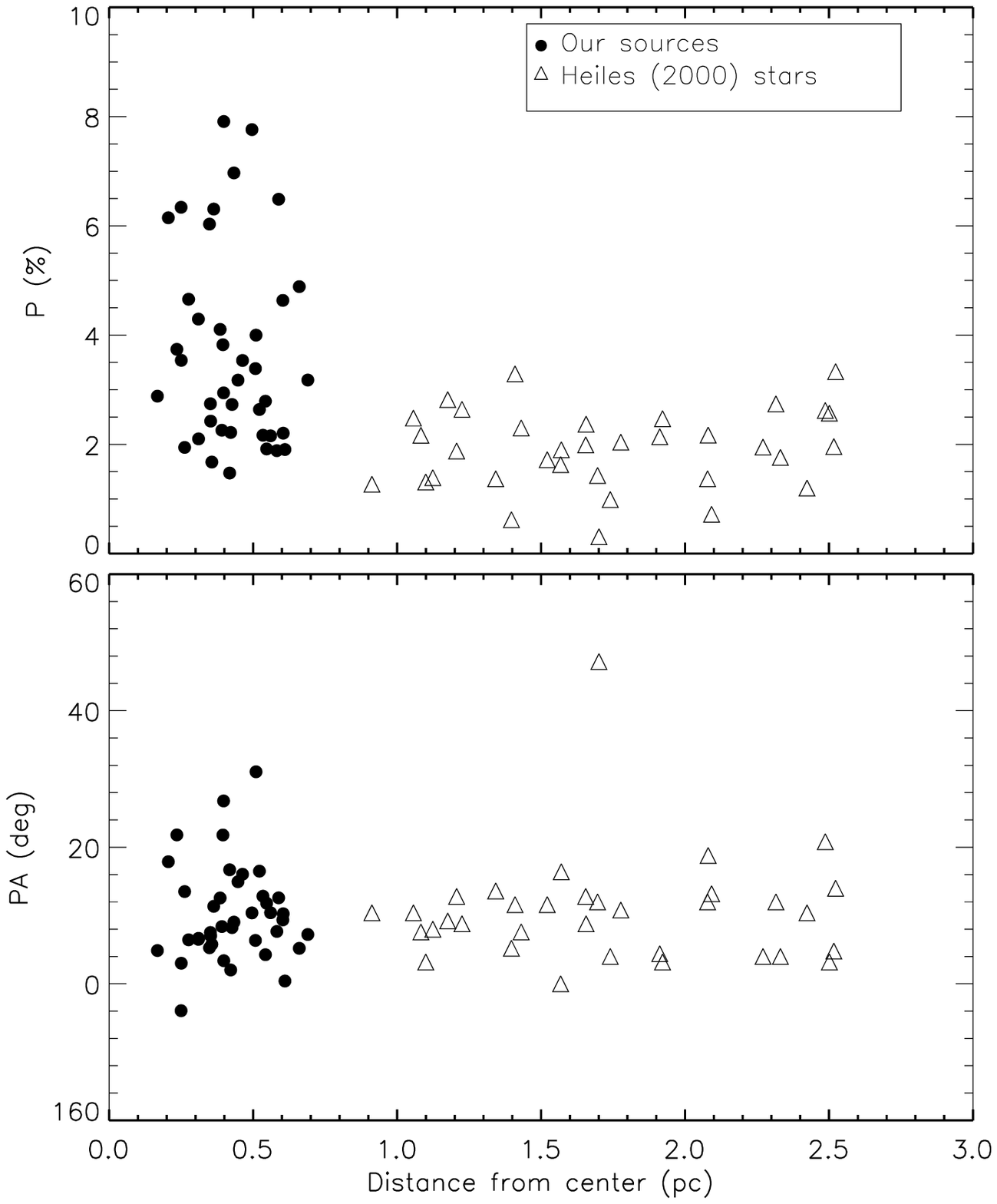}}
\caption{Variation of P and $\theta_{p}$ of the sources against their distances from the center of the core in L1521F. The stars we observed are shown as filled circles and those from the \citet{2000AJ....119..923H} catalogue, chosen within a circular region of 1$\degree$ around L1521F, are shown as open triangles. The location of the peak emission in CO towards L1521F \citep{2004A&A...420..957C} is chosen as the core center.}\label{Fig:PAdistCentl1521f}
\end{figure}
%#########################################################################################

In Fig. \ref{Fig:dss_iras1521f}, we show the polarisation vectors on the IRAS 100$\mu$m (the cloud morphology is not clearly discernible in WISE 12$\mu$m) image of the field containing L1521F and the position of the VeLLO, L1521F-IRS. The P versus $\theta_{p}$ plot obtained for the 40 stars, projected on L1521F, are presented in Fig. \ref{Fig:PvsPAl1521f}. The histogram of the $\theta_{p}$ values is also shown in the same plot. The mean value of P with its corresponding error is found to be 4.5$\pm$1.8\%. The mean value and the standard deviation obtained from a Gaussian fit to the $\theta_{p}$ are found to be 22$\degree$ and 13$\degree$. L1521F-IRS, detected by the SST \citep{2006ApJ...649L..37B}, shows scattered radiation with a bipolar morphology indicating the presence of an outflow associated with this class 0 source \citep{2009ApJ...696.1918T}. Recently, using $^{12}$CO (2-1) high resolution observations, \citet{2013ApJ...774...20T} for the first time, spatially resolved a compact but poorly collimated molecular outflow associated with L1521F-IRS. The outflow axis is roughly aligned at a position angle of 75$\degree$ \citep{2009ApJ...696.1918T,  2013ApJ...774...20T}. The outflow direction is shown with a solid line in Fig. \ref{Fig:dss_iras1521f} and with a broken line in Fig. \ref{Fig:PvsPAl1521f}.

From the \citet{2000AJ....119..923H} catalogue we obtained polarisation measurements for normal stars (excluding the peculiar, emission stars and those in binary system based on the information provided in the SIMBAD) that are located within a circular region of 1$\degree$ radius about L1521F (see Fig. \ref{Fig:PvsPAl1521f}). We selected only those measurements for which the P/$\sigma_{P}$ $\geq$ 2. The mean and the standard deviation of the P and $\theta_{p}$ for stars from the \cite{2000AJ....119..923H} catalogue are found to be 1.9$\pm$0.8\% and 26$\pm$20$\degree$. In Fig. \ref{Fig:heilesL1521f}, we show the polarisation vectors of stars observed by us and those from the \citet{2000AJ....119..923H} catalogue overplotted on IRAS 100$\mu$m image containing L1521F. In Fig. \ref{Fig:PAdistCentl1521f}, the distribution of P and $\theta_{p}$ of the stars observed by us and those obtained from the \citet{2000AJ....119..923H} catalogue are plotted as a function of the distance from center of L1521F; the location of the peak emission in CO towards L1521F \citep{2004A&A...420..957C} is chosen as the core center. The value of P is found to increase from the intercloud region to the periphery of the L1521F where we made polarisation observations. The comparable values of the mean of $\theta_{p}$ of the sources observed by us and that of the stars from the \citet{2000AJ....119..923H} catalogue (22$\degree$ and 26$\degree$, respectively) suggest that the envelope magnetic field is anchored to the magnetic field in the intercloud region. The orientation of the minor axis of L1521F estimated by \citet{2008A&A...487..993K} based on the morphology of the cloud in MAMBO\footnote{Max-Planck Millimeter Bolometer array at the IRAM 30-metre telescope} map in 1.2 mm is found to be 82$\degree$. The dashed line in Fig. \ref{Fig:dss_iras1521f} represents the Galactic plane at the latitude of the cloud oriented at an angle of 135$\degree$.

The velocity structure of L1521F was investigated by \citet[][]{2004A&A...420..957C} using N$_{2}$H$^{+}$ and N$_{2}$D$^{+}$ lines and \citet[][]{2004ApJ...601..962S} using CCS and N$_{2}$H$^{+}$ lines. The velocity structure of L1521F is found to be very complex with portions of the core showing rapid change in the gradient direction. Assuming that the velocity gradients are due to rotation, the results obtained by \citet[][]{2004ApJ...601..962S} suggest that the rotation of the outer envelope of the core (size $\lesssim$ 0.08 pc $\sim$ 16,000 AU) traced by the CCS components and that of the central compact region of the core (size $\lesssim$ 0.03 pc $\sim$ 6000 AU) traced by N$_{2}$H$^{+}$ have different axes with almost opposite senses of rotation. The global velocity gradient in L1521F traced by CCS components is from east to west and is almost along the major axis of the cloud.

\subsection{L328}

%#########################################################################################
\begin{figure}[!h]
\centering
\resizebox{9cm}{8cm}{\includegraphics{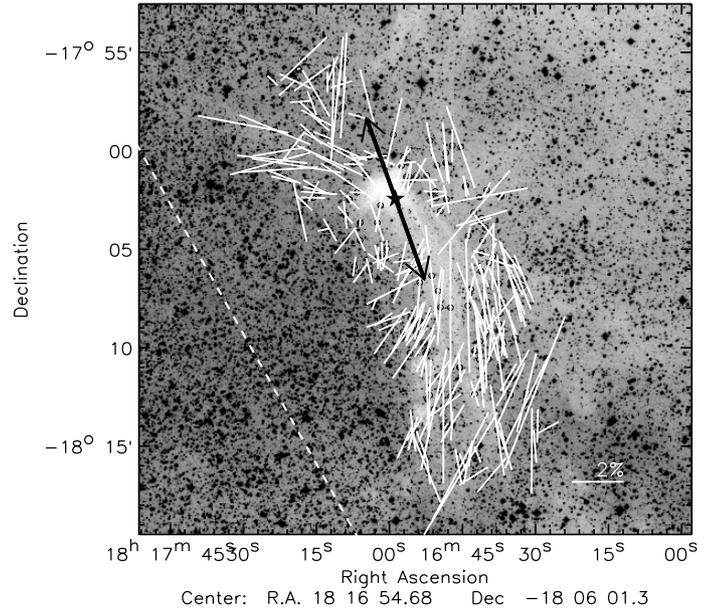}}
\caption{Optical polarisation vectors are over-plotted on $27^{\prime}$ R-band DSS image containing L328. The position of L328-IRS is shown by a black star. The outflow orientation is shown as a solid line with double arrow head and the Galactic plane at the latitude of the cloud is represented by a broken line. A vector corresponding to 2\% polarisation is shown as reference.}\label{Fig:dss_wise328_ls}
\end{figure}
%#########################################################################################
\begin{figure}
\centering
\resizebox{8cm}{8cm}{\includegraphics{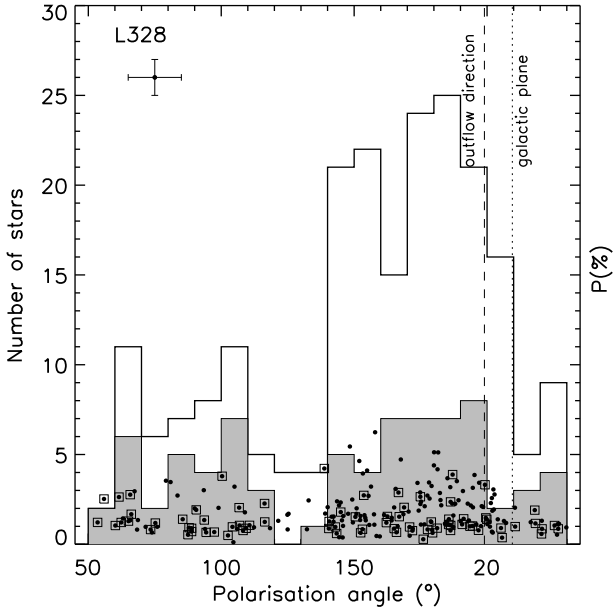}}
\caption{P versus $\theta_{p}$ for the stars observed towards L328. The histogram of the $\theta_{p}$ with binsize 10$\degree$ are also shown. Open histogram corresponds to the $\theta_{p}$ observed on entire cloud. Filled histogram is for the $\theta_{p}$ of the stars in $10^{\prime}$ region around L328-IRS. Filled circles represent the values estimated for entire cloud and the filled circles enclosed within the squares show the values measures around L328-IRS. The CO outflow direction associated with L328-IRS is shown as a dashed line. The position of the Galactic plane is shown as a dotted line. A typical error-bar is shown in the figure.}\label{Fig:PvsPAl328}
\end{figure}
%#########################################################################################
\begin{figure}
\resizebox{9cm}{8cm}{\includegraphics{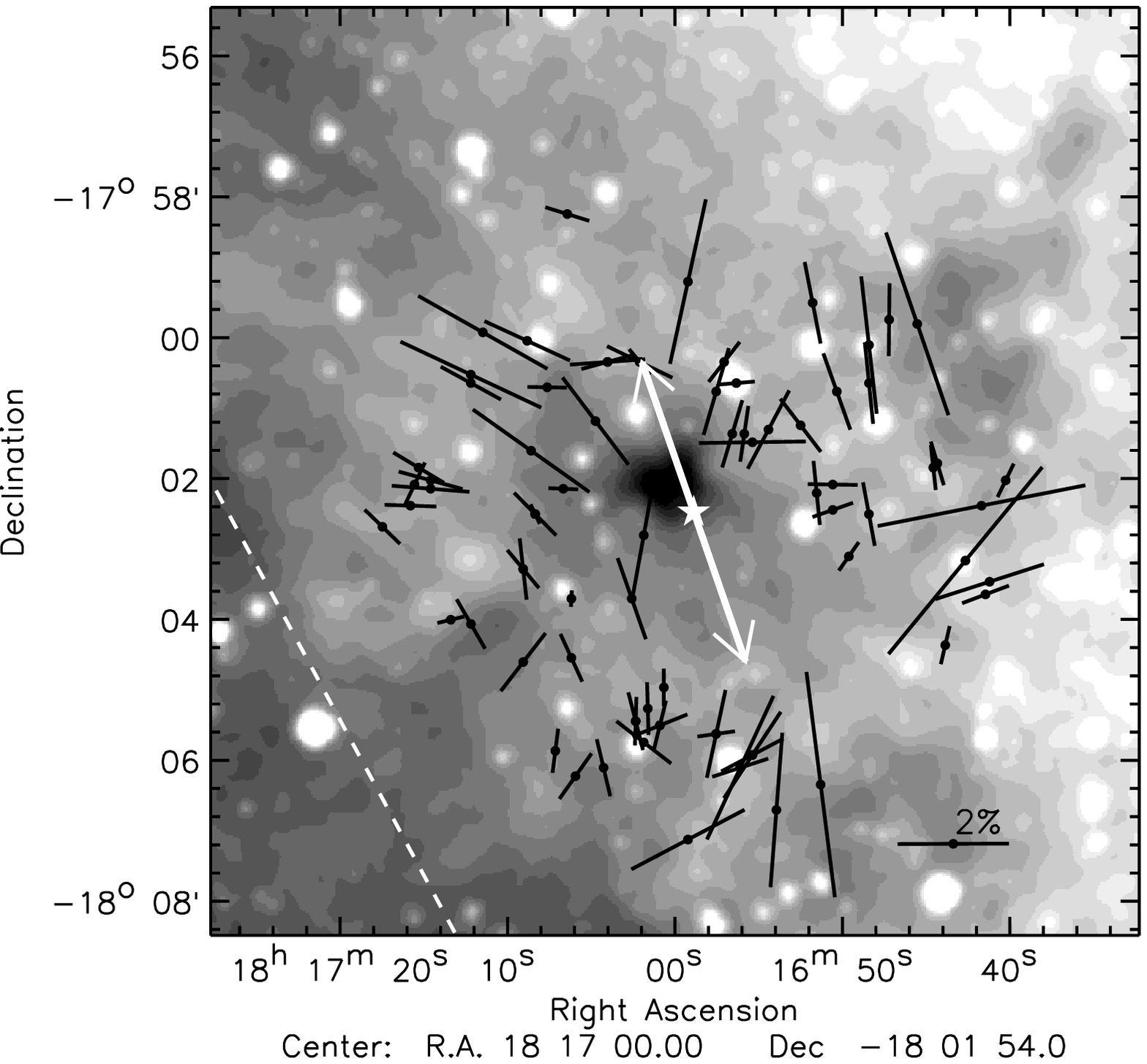}}
\caption{Optical polarisation vectors over-plotted on $0.22^\circ\times0.22^\circ$ WISE 12$\mu$m images of the field containing L328. The broken line shows the direction of the Galactic plane at the latitude of the cloud. The source L328-IRS is identified by a star. The molecular outflow direction is shown as a solid line with double arrow head. The vector with 2\% polarisation is shown as a reference.}\label{Fig:dss_wise328}
\end{figure}
%#########################################################################################

Towards L328, we obtained polarisation measurements of 215 stars that are projected on the entire cloud. The cloud shows a head-tail morphology in the R-band DSS image as shown in Fig. \ref{Fig:dss_wise328_ls}. The L328 core is located at the head and the tail extends $\sim15^{\prime}$ towards the southern direction. The polarisation vectors are found to be well ordered and seem to follow the large-scale structure of the cloud inferred from the DSS R-band image (Fig. \ref{Fig:dss_wise328_ls}). The distribution of $\theta_{p}$ for the entire cloud (open histograms) is presented in Fig. \ref{Fig:PvsPAl328}. The plot shows two dominant components of $\theta_{p}$ with broad distributions and without clear peaks. In the same plot we also show the P versus $\theta_P$ (closed circles) for the entire cloud. The mean value of the $P$ is found to be 1.8\%. The mean values of $\theta_{p}$ for the the two dominant components are found to be 87$\degree$ and 176$\degree$. The 87$\degree$ component is dominant towards the head region of L328 and the 176$\degree$ component is dominant towards the tail parts of the cloud. Star 31 is identified as an emission line star \citep{1981A&AS...44..387M} with observed values of P and $\theta_{p}$ as 1.2$\pm$0.2\% and 162$\pm$5$\degree$. The mean value of 176$\degree$ is found unchanged after the exclusion of star 31.

A candidate proto-brown dwarf \citep{2009ApJ...693.1290L, 2013ApJ...777...50L}, L328-IRS, of very low luminosity was detected in L328 using the SST. The position of L328-IRS is identified in Fig. \ref{Fig:dss_wise328_ls}. In order to estimate the mean magnetic field direction in the vicinity of the L328-IRS, we choose a circular region of 10$^{\prime}$ about it. The polarisation vectors of 77 stars from this region are shown on the WISE 12$\mu$m images (Fig. \ref{Fig:dss_wise328}). The L328 core is clearly visible as absorption in WISE 12$\mu$m image. The distribution of  $\theta_{p}$ of the stars from this region, shown in Fig. \ref{Fig:PvsPAl328}, is similar to the one shown for the entire cloud. The same plot also shows the P versus $\theta_P$ for the stars from the above region. The mean values of the two components of the distributions for the stars around L328 core are 87$\degree$ and 1$\degree$. In the vicinity of the L328-IRS, we have taken the average value (44$\degree$) of these two components as the orientation of the mean magnetic field. \citet{2009ApJ...693.1290L} showed that a velocity gradient from north-east (6.3 km$s^{-1}$) to south-west (6.7 km$s^{-1}$) is present in the L328 core. The velocity gradient seems to follow the cloud structure which is possibly affected by the presence of an external ionising agent. Recently we made near-IR polarimetry of the core region of L328 and the preliminary results show that the stars located close to L328-IRS have a well-ordered distribution approximately at an angle of $\sim45\degree$ similar to the average value of 44$\degree$ obtained here. A detailed study of L328 will be presented in a subsequent paper in which we will present the results of near-IR polarimetry of the core region.

The near-IR images presented by \citet{2009ApJ...693.1290L} show nebulosity around L328-IRS which they attribute to an outflow cavity, strongly supporting the possible existence of outflow activity associated with L328-IRS. From the CO bipolar outflow associated with L328-IRS \citep{2013ApJ...777...50L}, we estimated a position angle of $\sim20\degree$ for the outflow direction. L328 is shown to consist of three sub-cores called 1, 2, and 3 \citep{2009ApJ...693.1290L}. The VeLLO, L328-IRS, is found to be associated with the sub-core 2. The outflow position angle is shown in Fig. \ref{Fig:dss_wise328}. The Galactic plane at an angle of 29$\degree$, at the latitude of the cloud, is drawn as a dashed line in Fig. \ref{Fig:dss_wise328_ls}.

\subsection{L673-7}

%##############################################################################################
\begin{figure}
\centering
\resizebox{8.5cm}{8.5cm}{\includegraphics{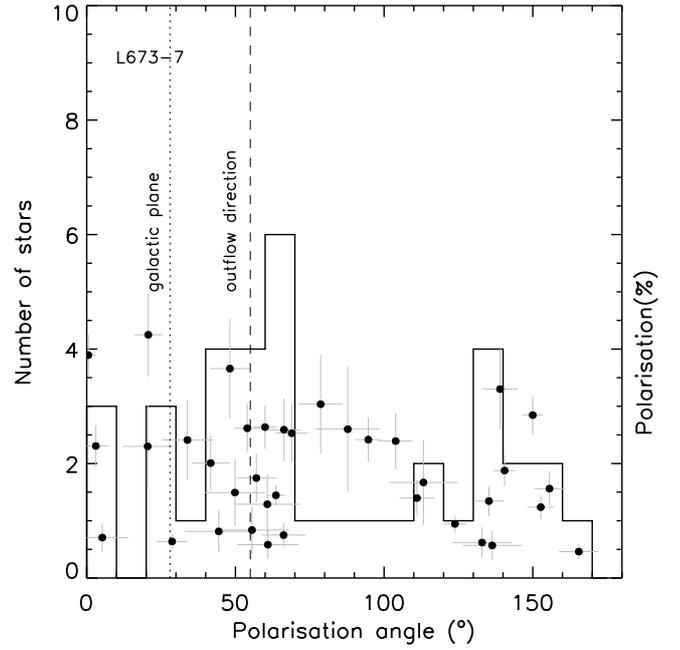}}
\caption{P versus $\theta_{p}$ for the stars observed towards L673-7. The histogram of the $\theta_{p}$ with binsize 10$\degree$ is also shown. The molecular outflow direction associated with L673-7-IRS is shown as a dashed line. The direction of the Galactic plane at the latitude of the cloud is shown as a dotted line.}\label{Fig:PvsPAl673}
\end{figure}
%##############################################################################################
\begin{figure}
\resizebox{8.5cm}{8.5cm}{\includegraphics{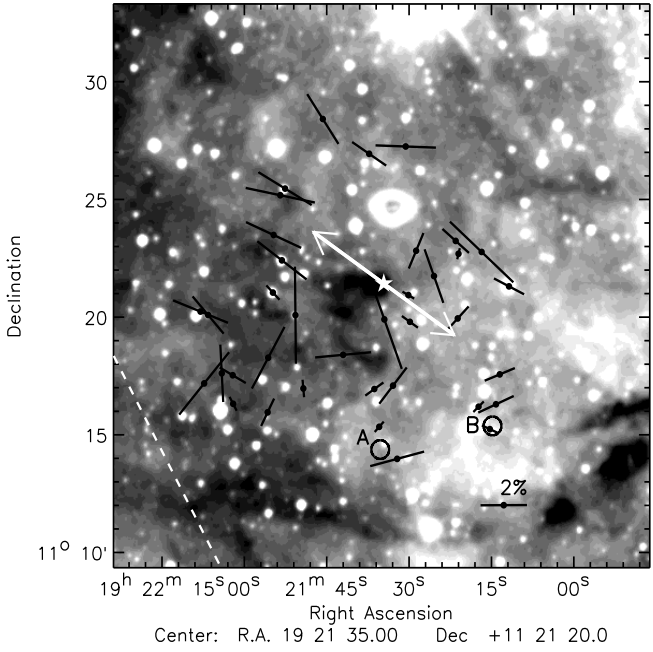}}
\caption{Polarisation vectors over-plotted on $0.4^\circ$ $\times$ $0.4^\circ$ WISE 12$\mu$m image of the field containing L673-7. The source L673-7-IRS is identified by a star. The molecular outflow direction is shown as a solid line with double arrow head. The broken line shows the direction of the Galactic plane at the latitude of the cloud. The vector with 2\% polarisation is shown as reference. The positions of the SRAO observations are identified as circles A and B.}\label{Fig:dss_wise673}
\end{figure}
%(in a region of $\sim$1.1 pc)
%##############################################################################################
\begin{figure}
\resizebox{8.5cm}{8.5cm}{\includegraphics{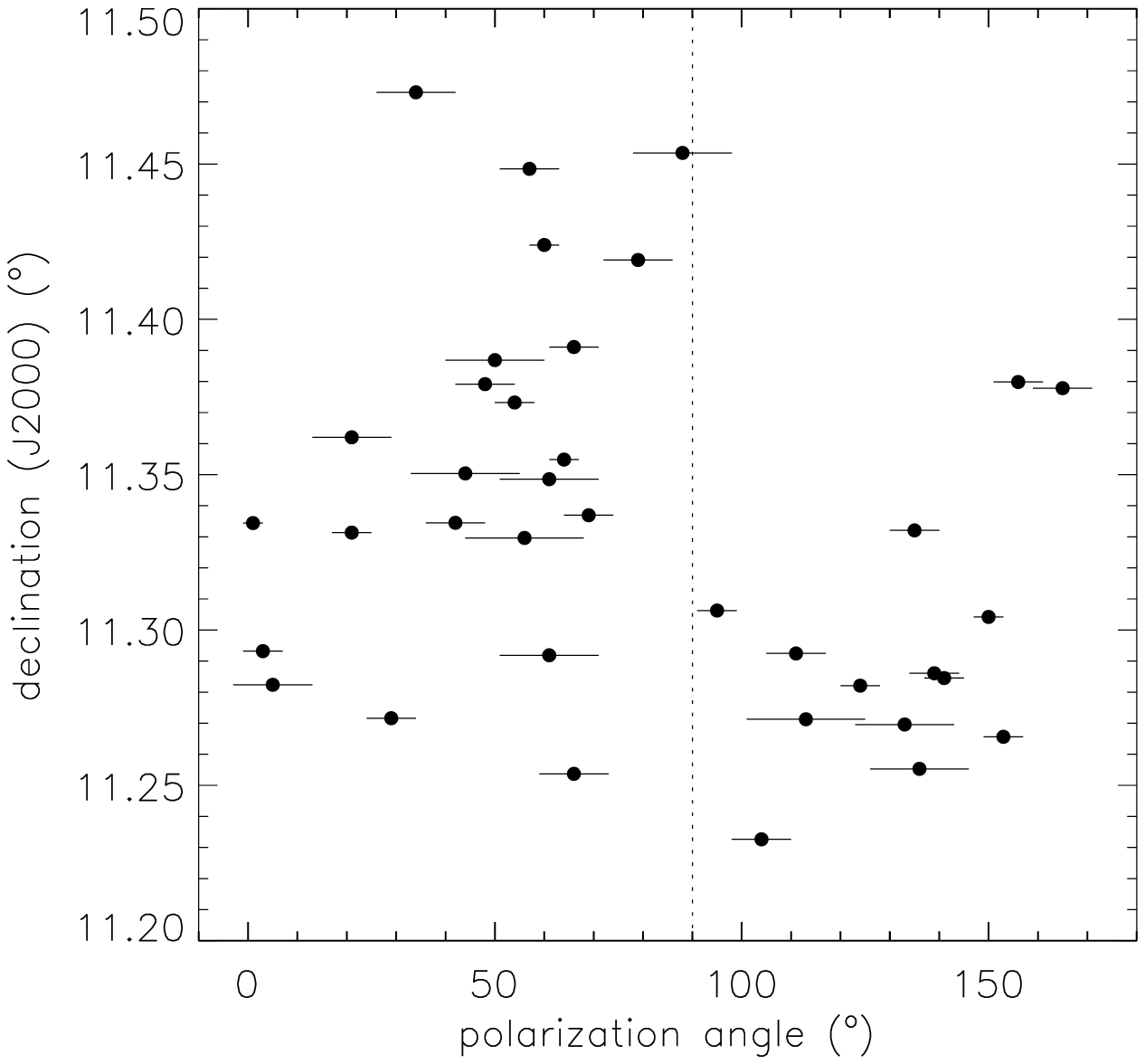}}
\caption{The distribution of $\theta_{p}$ towards L673-7 with respect to the declination. The dotted line is drawn to show the two populations of sources that are present towards L673-7.}\label{Fig:l673pol_dec}
\end{figure}
%##############################################################################################

L673-7 is one among the eleven cores identified by \citet{1999ApJS..123..233L} towards the L673 complex. \citet{2010ApJ...721..995D}, based on the data obtained from the SST, reported the discovery of a VeLLO, L673-7-IRS, an embedded Class 0 protostar driving a molecular outflow. The $\theta_{out}$ is estimated to be $\sim55\degree$ with respect to the north. This is inferred based on the prominent redshifted and blueshifted peaks identified in the CO molecular line observations \citep{2010ApJ...721..995D}. 

In Fig. \ref{Fig:PvsPAl673}, we present the $P$ versus $\theta_{p}$ plot for the 37 stars we observed towards L673-7. The histogram of the $\theta_{p}$ is also presented in the same plot. The polarisation vectors are over-plotted on the WISE 12$\mu$m images as shown in Fig. \ref{Fig:dss_wise673}. The mean values of P and $\theta_{p}$ are found to be 1.9$\pm$1.0\% and 79$\pm$48$\degree$. The $\theta_{p}$ values show a very scattered distribution. The polarisation vectors to the northern parts of the core seem to be more ordered than those distributed to the southern parts. In Fig. \ref{Fig:l673pol_dec}, we show the distribution of $\theta_{p}$ with respect to the declination. We noticed the presence of two populations of sources in the plot. The population in the northern part predominantly shows $\theta_{p}$ less than 90$\degree$, and the other population towards the southern parts shows a more mixed distribution of $\theta_{p}$.

The two populations of sources seen towards L673-7 could be due the presence of two distinct clouds in the line of sight. In order to examine this possibility, we looked at the CO (J=2-1) molecular line data obtained by \citet[in preparation;][]{inpreparation} towards two locations in the southern region of L673-7 where position angles are different from the one corresponding to the envelope of the cloud. The data were taken using the Seoul Radio Astronomy Observatory (SRAO) 6~m radio telescope. The positions of the SRAO observations are identified using circles A and B in Fig. \ref{Fig:dss_wise673}. The offsets of A and B are ($0^{\prime\prime}$,$-420^{\prime\prime}$) and ($-300^{\prime\prime}$,$-360^{\prime\prime}$) about L673-7. The spectra show very strong 51-53 km$s^{-1}$ component as well as the 7.4 km$s^{-1}$ component. The 7.4 km$s^{-1}$ component is associated with L673-7. In the vicinity of the L673-7 core, there is a bright 23 km$s^{-1}$ component which could be in the background of L673-7. However, this component seems to get weaker at the southern part of L673-7. Although observations using SRAO do not cover the whole area where polarisation data were obtained in L673-7, the spectra from the two observed locations clearly indicate that there are velocity components of 51-53 km$s^{-1}$ which seem to be widely distributed in the background of southern parts of L673-7. These components can cause the different magnetic field structure in the southern part of L673-7. Mapping of the whole southern parts of the cloud is required to ascertain whether the 51-53 km$s^{-1}$ components are the real cause for the additional observed magnetic field components in the southern area of L673-7.

The average value of the P with the standard deviation is found to be 1.9$\pm$0.9\%. The mean value of the polarisation $\theta_{p}$ of the stars from the northern region, considered as the mean magnetic field direction, is estimated to be 47$\pm$24$\degree$. The position angle of the minor axis of the cloud, obtained from the 1.2 mm MAMBO images, is found to be 0$\degree$ \citep{2008A&A...487..993K}. The Galactic plane at the latitude of the cloud is found to be $28\degree$ (see Fig. \ref{Fig:dss_wise673}).

\subsection{L1014}
%##############################################################################################
\begin{figure}
\resizebox{8cm}{8cm}{\includegraphics{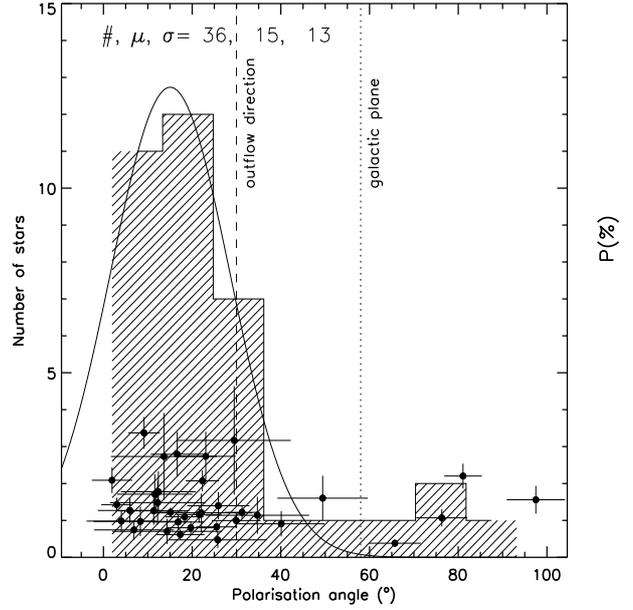}}
\caption{The P versus $\theta_{p}$ of the stars observed in the direction the of L1014. The histogram of the $\theta_{p}$ with binsize 10$\degree$ is also shown. The direction of the outflow is shown as dashed line. The direction of the Galactic plane at the latitude of the cloud is shown as a dotted line.}\label{Fig:PvsPAl1014}
\end{figure}
%##############################################################################################
\begin{figure}
\resizebox{8.4cm}{8cm}{\includegraphics{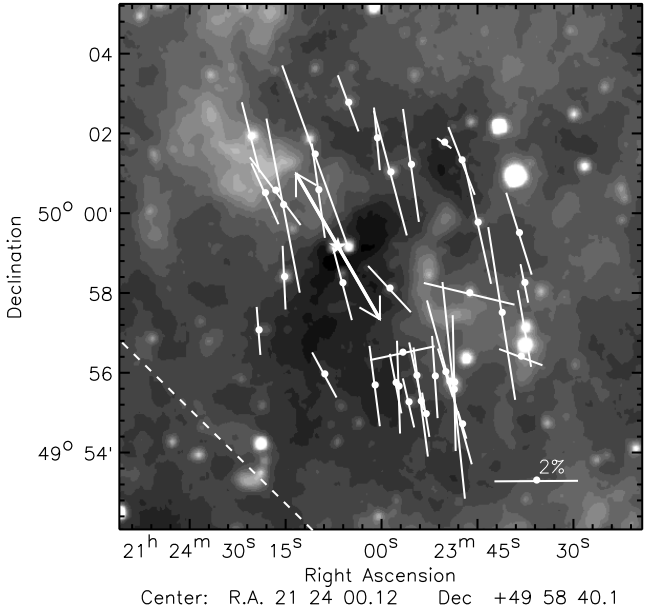}}
\caption{Optical polarisation vectors over-plotted on $0.22^\circ$ $\times$ $0.22^\circ$ WISE 12$\mu$m image of the field containing L1014. The source L1014-IRS is identified as a star. The broken line shows the direction of the Galactic plane at the latitude of the cloud. The molecular outflow direction is shown as a thick line with double arrow head. The vector with 2\% polarisation is shown as reference.}\label{Fig:dss_wise1014}
\end{figure}
%##############################################################################################

In Fig.\ref{Fig:PvsPAl1014}, we present the $P$ versus $\theta_{p}$ plot of the 38 stars observed by us towards L1014. In the same plot we show the histogram of the $\theta_{p}$ values. The results are plotted over the WISE 12$\mu$m images as shown in Fig. \ref{Fig:dss_wise1014}. L1014 is one of the first starless cores in which a VeLLO was detected by the SST \citep{2004ApJS..154..396Y}. The VeLLO, L1014-IRS, is found to be associated with a weak but compact molecular outflow along a direction of 30$\degree$ from the north \citep{2005ApJ...633L.129B}. Outflow direction is shown with a dashed line in Fig. \ref{Fig:PvsPAl1014}. The mean of the P with the standard deviation is estimated to be 1.9$\pm$0.8\%. A Gaussian fit to the $\theta_{p}$ yields a mean of $\sim15\pm$11$\degree$. As in the case of L673-7, L1014 dense core could also be seen in absorption of the mid-infrared background. The vectors are found to be distributed smoothly. The Galactic plane at the latitude of the cloud is found to be $57\degree$ (see Fig. \ref{Fig:dss_wise1014}). The stars 6 and 25 observed by us are identified as emission line sources by \citep{1999A&AS..134..255K}. The values of P and $\theta_{p}$ of these stars are 1.9$\pm$0.2\%, 1.2$\pm$0.2\% and 19$\pm$2$\degree$, 55$\pm$6$\degree$, respectively. The mean and standard deviation in P and $\theta_{p}$ of the sources observed towards L1014 after excluding stars 6 and 25 are found to be 1.5$\pm$0.8\% and 15$\pm$13$\degree$, respectively. The orientation of the minor axis of the cloud is at an angle of 70$\degree$ as obtained from the 1.2 mm MAMBO image of L1014 \citep{2008A&A...487..993K}.

\section{Discussion}\label{Discussion}

In Table \ref{tab:offsets}, we present the angular offsets between various properties like the envelope magnetic field ($\theta^{opt}_{B}$,  obtained from the optical polarisation measurements), core size magnetic field ($\theta^{sub}_{B}$, obtained from the submm polarisation measurements), minor axis ($\theta_{min}$) of the cloud cores and the outflow directions ($\theta_{out}$). The interpretations based on the angular offsets given in Table \ref{tab:offsets} should be taken cautiously as the calculations are made between two projected quantities. 

\subsection{Relative orientation between magnetic field and outflows}

The outflows are detected in all of the five cores studied here. In IRAM 04191, the position angle of the outflow is estimated using the CO molecular line observations. The outflow is found to be highly collimated \citep{2002A&A...393..927B}. However, the outflows detected in L1521F \citep{2013ApJ...774...20T}, L328 \citep{2013ApJ...777...50L}, and L673-7 \citep{2010ApJ...721..995D} are found to be poorly collimated with wide opening angles. The VeLLO L1014-IRS is found to be associated with a weak but compact molecular outflow \citep{2005ApJ...633L.129B}. The position angles of outflows from sources with such wide opening angles are difficult to measure and often results in uncertainty. For example, the position angles of molecular outflows obtained from CO(2-1) molecular line emission is found to be 28$\degree$ \citep{2002A&A...393..927B} and 30$\degree$ \citep{2005ApJ...633L.129B} in IRAM 04191 and L1014. These are the values we have considered for the present study, but the recent values of outflow position angles given by \citet{2014ApJ...783...29D} for these cores are 22$\degree$ and 45$\degree$. Therefore, there is always an uncertainty of $\sim$10-15$\degree$ in the estimation of outflow position angles.

The angular offsets between $\theta^{opt}_{B}$ and $\theta_{out}$ for IRAM 04191, L1521F, L328, L673-7 and L1014 are 84$\degree$, 53$\degree$, 24$\degree$, 8$\degree$, and 15$\degree$, respectively (given in Col. 7 of Table \ref{tab:offsets}). The mean value of the offsets for the five clouds is $\sim37\degree$. If we exclude IRAM 04191 which shows an almost perpendicular orientation, the mean value reduces to $\sim25\degree$. Based on the MHD simulations on a slowly rotating, magnetised molecular cloud core undergoing gravitational collapse, \citet{2004ApJ...616..266M} showed that the alignment of the outflow with a local magnetic field at $\sim10$ AU scale is independent of the magnetic field strengths considered in the simulations but on the cloud scale, the alignment depends on the magnetic field strengths assumed. The alignment is better when the assumed magnetic field is stronger. Studies conducted to test the relative alignment between the cloud scale magnetic field and outflows, for example, from T-Tauri stars in the Taurus molecular cloud \citep{2004A&A...425..973M} have found no correlation. \citet{2011ApJ...743...54T}, based on the results from optical polarisation observations of protostars with a range of ages, also found a random distribution for the offsets between magnetic field and outflow directions. However, statistical evidence of alignment was noticed when only class 0/I sources were considered. A possible explanation for the lack of alignment found between cloud scale magnetic field and the outflow could be that the more evolved T-Tauri stars might have injected more turbulence into their surroundings which could have scrambled any alignment that might have existed in the early evolutionary stages \citep{2013ApJ...770..151C}. 

The protostars associated with the cores studied in this work are all found to be in their early (class 0) stage of evolution. The momentum flux estimated for IRAM 04191-IRS \citep{1999ApJ...513L..57A} is the highest among the VeLLOs. This value comes in between class 0 and I sources of the Perseus molecular cloud studied by \citet{2010MNRAS.408.1516C}. The momentum flux estimated for L1521F-IRS \citep{2013ApJ...774...20T}, L673-7-IRS \citep{2012AJ....144..115S}, and L1014-IRS \citep{2005ApJ...633L.129B} are found be the lowest among the protostars and that of L328-IRS \citep{2013ApJ...777...50L} is the lowest among the VeLLOs. This implies that the outflow from VeLLOs are less powerful than those of any other protostars. Thus the VeLLOs are expected to inject the lowest possible turbulence into their surrounding environment enabling the clouds to retain their initial magnetic field orientation. The offset between outflow and the cloud scale magnetic field directions in three (L328, L673-7, and L1014) out of five clouds studied here are found to be aligned within $\sim25\degree$. In IRAM 04191, the outflow is oriented almost perpendicular to the direction of the cloud scale magnetic field which is further discussed in section \ref{low_high}.

In order to examine the effect of magnetic field strength on the alignment between the field and the outflow, we estimated the strength of the plane-of-the-sky component of the magnetic field using the Chandrasekhar-Fermi (CF) relation ($B_{POS}=9.3\sqrt{n(H_{2})}\delta v/\delta\theta$) \citep{1953ApJ...118..113C, 2001ApJ...546..980O, 2005mpge.conf..103C}. Here $n(H_{2})$ represents the volume density of the clouds, $\delta v$ is the velocity dispersion, and $\delta\theta$ is the dispersion in $\theta_{P}$ corrected by the uncertainty in $\theta_{P}$. We adopted the $^{13}$CO line width ($\Delta v$) towards these clouds from \citet[in preparation;][]{inpreparation}. To estimate the magnetic field strength, we have used the velocity dispersion ($\delta v$) calculated from the line width ($\Delta v/(8ln2)^{1/2}$). We have assumed a typical volume density of $\simeq 3\times 10^{3} cm^{-3}$ for the regions up to which optical observations were made in these clouds. The position angle dispersion ($\delta\theta$) used to calculate the magnetic field strength has been estimated from the standard deviation obtained from a Gaussian fit to the $\theta_{P}$ in these clouds (see Table \ref{tab:offsets}) and corrected for the position angle uncertainty \citep{2001ApJ...561..864L, 2010ApJ...723..146F}. Assuming these values, we estimated a magnetic field strength of $\sim28\mu$G, $\sim24\mu$G, $\sim15\mu$G and $\sim20\mu$G for IRAM 04191, L1521F, L328, and L1014, respectively. Because of high dispersion in $\theta_{P}$ towards L673-7, the basic assumption of the CF method (i.e. $\delta\theta>25\degree$) breaks down and hence its use would render an inaccurate magnetic field strength. The magnetic field strength of the inner parts of IRAM 04191 was also not calculated because of the high dispersion in $\theta_{P}$ obtained from SCUPOL. Adopting the uncertainties in $\theta_{P}$ and dispersion velocity, the typical uncertainty ($\sigma B_{POS}$) in the magnetic field strength is estimated to be $\sim0.5B_{POS}$. The line-of-sight magnetic field strengths given by \citet{2008ApJ...680..457T} and \citet{2010ApJ...725..466C}, are found to be 3.3$\pm3.5\mu$G and -1.4$\pm4.4\mu$G in IRAM 04191 and L1521F.. At the cloud scale, the dependence of alignment between the magnetic field and the outflow direction on the magnetic field strengths is not very apparent in the clouds studied here.

The offset between the inner magnetic field (inferred from SCUPOL data in $\sim1000 - 10000$ AU scale) and the direction of outflow in IRAM 04191 is found to be $\sim16\degree$. Recently, \citet{2014ApJS..213...13H}, based on results from $\lambda$1.3~mm dust polarisation observations of protostellar cores found no correlation between the mean magnetic field directions at $\sim$1000 AU scales and the protostellar outflow axes. They found that the cores with high polarisation (P$\gtrsim$3\%) showed magnetic fields consistent from large scales (0.1 pc) to small scales (0.01 pc) and the magnetic field and outflows in them are randomly oriented. In the low polarisation (P$\lesssim$3\%) sources, the magnetic fields at large scales are not consistent with the small scales and the outflows are oriented preferentially perpendicular to the magnetic field directions. In IRAM 04191, the outer magnetic field is perpendicular to the inner magnetic field (see sect. 4.2). The mean value of P (from SCUPOL data) around IRAM 04191 is estimated to be $7\pm4$\% showing relatively high polarisation. The outflow is nearly parallel to the inner magnetic field orientation differing from the statistical conclusion made by \citet{2014ApJS..213...13H} for sources showing high polarisation values. For a carefully chosen sample of seven isolated, nearby, low-mass cores containing class 0 sources, \cite{2011ApJ...732...97D} and \citet{2013ApJ...770..151C}, based on 350$\mu$m polarisation observations, found evidence of a positive correlation between core magnetic field direction and bipolar outflow axis. It would be interesting to carry out submillimetre polarisation observations of cores with VeLLOs to understand the relative orientation between the outflow direction and the core scale magnetic field in them.

\subsection{Relation between magnetic fields in the low and high density regions}\label{low_high}

Of the five cloud cores studied here, only IRAM 04191 has the submillimetre observations available to compare the relationship between the envelope (inferred from the optical polarisation data) and the inner magnetic field (inferred from the submillimetre polarisation data). The angular offset between the envelope ($\sim0.2-0.7$ pc) and the inner magnetic field in IRAM 04191 is found to be 68$\degree$. If the formation of the core is mediated by the magnetic field then the field inside the core is expected to be inherited from the inter cloud region. In such a scenario, the magnetic field lines from the envelope are expected to correlate with the field lines of the inner parts of the core. \citet{2009ApJ...704..891L} showed that there was a significant correlation between the magnetic field of the cloud cores inferred from the submillimetre polarimetric data and that of the inter cloud region. Then, the almost perpendicular orientations of the magnetic field at the envelope of IRAM 04191 with respect to that of the inner core looks anomalous. 

Recently in the starless core, L183, based on the near-IR and 850$\mu$m polarimetry, \citet{2012ApJ...748...18C} found that the inner and outer magnetic fields are oriented almost perpendicular to each other. They suggested that the reason for this offset could be either due to the decoupling of the core and envelope magnetic field directions for some intermediate cloud radii, or coherent twisting of the field with radius. \textit{In IRAM 04191, is there a possibility that the magnetic field lines were modified after the formation of the core?} Interestingly, the cloud shows a wind-blown morphology in the 12$\mu$m image (see Fig. \ref{Fig:dss_wiseIRAM}). The wind-blown morphology of the clouds are common towards the regions that are in the vicinity of O- or early B-type stars and towards clusters with one or more luminous sources. The dynamical evolution of molecular clouds in the vicinity of luminous sources have been studied numerically by a number of authors \citep[e.g. ][]{1989ApJ...346..735B, 1990ApJ...354..529B, 1994A&A...289..559L, 2001MNRAS.327..788W, 2006MNRAS.369..143M, 2009MNRAS.398..157H, 2010MNRAS.403..714M, 2011MNRAS.412.2079M}. Some of these studies have included magnetic fields of varying orientations and strengths to look at the evolution of the magnetic field structure and the cloud morphology under the influence of the external source(s) \citep[e.g. ][]{2011MNRAS.412.2079M}. The results obtained from such studies have shown that in the case of weak and medium field strengths an initially perpendicular field is swept into complete or partial alignment with the pillars or the head-tail structure of cometary globules during their dynamical evolution. Such scenarios have been observed in e.g. ESO 210-6A \citep{1987ApJ...319..842H}, CG 22 \citep{1996MNRAS.279.1191S}, CG 30 \citep{1999MNRAS.308...40B}, CG 12 \citep{2004MNRAS.348...83B},  M16 \citep{2007PASJ...59..507S}, LBN 437 \citep{2013MNRAS.432.1502S}. 

In IRAM 04191, if we consider the inner magnetic field to be the original field direction, then the external agent which is responsible for the wind-blown morphology of the cloud could have dragged the outer magnetic field to align with it. However, we could not find a potential source (or sources) in the vicinity of IRAM 04191 that is responsible for the present morphology of the cloud. 

\subsection{Relative orientation between magnetic field, minor axis of the cores, and outflows}

Out of the five cores presented here, the minor axis position angles for four of them namely, IRAM 04191, L1521F, L673-7, and L1014 (see Table \ref{tab:parameters}) are available in the literature. These values of minor axes are inferred based on the 850$\mu$m and 1.2~mm dust continuum emission maps available in the literature. The angular offset between $\theta^{opt}_{B}$ and $\theta_{min}$ for IRAM 04191, L1521F, L673-7, and L1014 are 82$\degree$, 60$\degree$, 47$\degree$, and 55$\degree$, respectively (listed in Col. 5 of Table \ref{tab:offsets}). The mean value of these offsets for IRAM 04191, L1521F, L673-7, and L1014 is found to be $\sim60\degree$. Thus, the minor axes of the cores is not aligned with the envelope magnetic field. This is inconsistent with the magnetically dominated star formation models. \citet{2000ApJ...540L.103B}, considering the clouds as triaxial bodies, showed that the average offset typically falls in the range of 10$\degree-30\degree$ when viewed from a random set of viewing angles. Therefore, the offset of $\sim60\degree$ between $\theta^{opt}_{B}$ and $\theta_{min}$ cannot even be explained as due to the projection effect. The minor axes position angles ($\theta^{opt}_{min}$) of L1521F, L328, and L1014 inferred by the method of visually fitting the ellipse to these optically selected cores are found to be 35$\degree$, 43$\degree$, and 32$\degree$, respectively \citep{1999ApJS..123..233L}. The values of the angular offset between $\theta^{opt}_{B}$ and $\theta^{opt}_{min}$ for L1521F, L328, and L1014 are 13$\degree$, 1$\degree$, and 17$\degree$, respectively. The mean value of the offsets for these cores is found to be $\sim10\degree$. Thus, the outer magnetic fields are found to be almost parallel to the minor axes inferred from the optically fitted ellipse to the cores than the minor axes inferred from the submillimetre continuum emission maps.

%============================================================
The projected magnetic field orientation of the core in IRAM 04191, inferred from the submillimetre polarimetry, on the other hand, is found to be parallel to the projected position angle of the minor axis (angular offset$\sim14\degree$; Col. 6 of Table \ref{tab:offsets}). \citet{2009MNRAS.398..394W} presented the offsets between the inner magnetic field and minor axis of the five cores observed previously, namely L1544, L183, L43, L1498, and L1517B, together with CB\,3 and CB\,246. Among them, CB\,3 is a star forming core. The weighted mean value of the offset between the inner magnetic field and the minor axis of the above five starless cores is found to be $\sim30\degree$ and that for CB\,3 (a star forming core) is $\sim40\degree$ \citep{2009MNRAS.398..394W}.

 In Col. 9 of the Table \ref{tab:offsets}, we present the angular offset between the projected position angle of the minor axis of the clouds inferred from submillimetre continuum emission and the outflows from the protostars. The mean value of the offsets between the minor axes of the four clouds (IRAM 04191, L1521F, L673-7, and L1014) and the outflow directions is found to be $\sim26\degree$. \citet{2013ApJ...770..151C} presented the 350 $\mu$m polarisation observations of four low-mass cores containing Class 0 protostars: L483, L1157, L1448-IRS2, and Serp-FIR1 to test the magnetically regulated models for core-collapse. They found a tight correlation between minor axes and outflow  direction. Based on this result they concluded that the inclination angle of the outflows could be used as the proxy of the minor axis inclination angle.

\subsection{Relative orientation between magnetic field and the Galactic plane}

Based on the observations of six giant molecular cloud complexes in the M33 galaxy, the magnetic field lines are shown to be aligned with the spiral arms suggesting that the large-scale field in M33 anchors the cloud \citep{2011Natur.479..499L}. The inter-cloud fields are also shown to correlate with the field orientations at much smaller scales \citep{2009ApJ...704..891L}. Therefore in a magnetic field mediated star formation scenario we do expect a coupling between the Galactic plane, the minor axis, and the outflow directions unless the turbulence and rotation of a cloud may randomise the orientation of the magnetic field in them. In a number of globules and dark clouds, the mean magnetic field pattern is found to be coupled with the Galactic plane by being more parallel to it \citep[e.g. ][]{1990AJ.....99..638K, 1995ApJ...445..269K}. However, it was noticed in a few cases that the fields are decoupled with the Galactic plane \citep{1987ApJ...319..842H, 1990ApJ...359..363G}. In five clouds studied here, the mean value of the offset angles between the Galactic plane and the projected outer magnetic field directions is found to be 34$\degree$. The inner magnetic field in IRAM 04191 is found to be almost perpendicular to the Galactic plane. The mean offset angle between the minor axis of the cloud cores and the Galactic plane is found to be $\sim40\degree$. If we exclude the two cloud cores associated with the Taurus star forming region namely IRAM 04191 and L1521F, this becomes $\sim20\degree$. 

\subsection{Polarisation efficiency}

%================================================================
\begin{figure}
\resizebox{8cm}{11cm}{\includegraphics{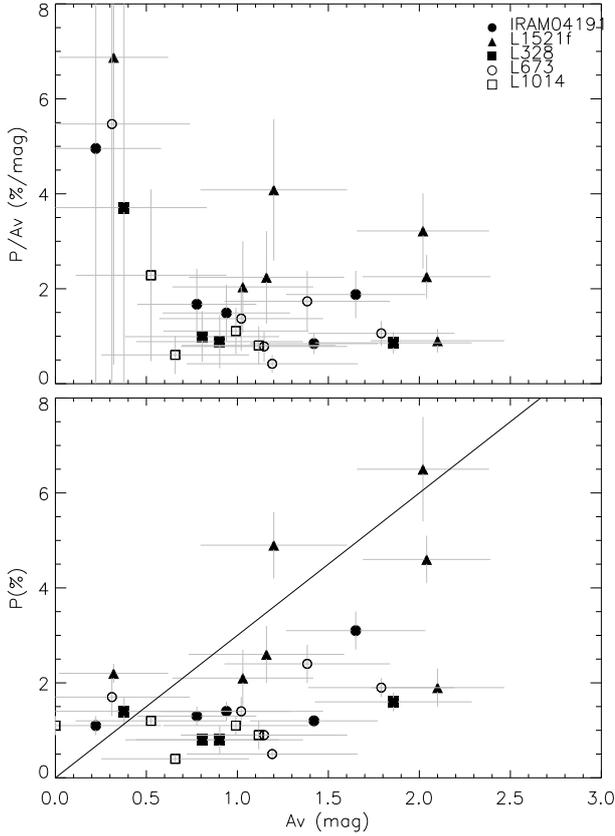}}
\caption{Upper panel: Polarisation efficiency (P/$A_{V}$) versus $A_{V}$ for a number of stars observed towards the five clouds studied in this work (shown with different symbols). Lower panel: P versus $A_{V}$ of the same stars in all five clouds. The solid line represents the observational upper limit of P/$A_{V}$ = 3\%/$mag^{-1}$.}\label{Fig:poleff}
\end{figure}
%================================================================

The polarisation efficiency of the dust grains towards a particular direction/line of sight is defined as the degree of polarisation produced for a given amount of extinction. Mie calculations for infinite cylindrical particles with dielectric optical properties that are perfectly aligned with their long axes parallel to each other and perpendicular to the line of sight place a theoretical upper limit on the polarisation efficiency of the grains that is because of selective extinction. This upper limit is found to be P/$A_{V}$ $\leq$ 14\% $mag^{-1}$ \citep{2003dge..conf.....W}. The observational upper limit on  P/$A_{V}$ is found to be 3\% $mag^{-1}$ \citep{2003dge..conf.....W}, however, a factor of four lower than the predicted value for the ideal scenario. Values of $A_{V}$ towards these five clouds studied here, are estimated using the 2MASS data,  and following the steps given in \citet{2008Ap&SS.315..215M}. The $A_{V}$ values could be measured only for a small number of stars in these clouds.

 Upper panel in Fig. \ref{Fig:poleff} plots the polarisation efficiency against the $A_{V}$ for a number of stars observed towards these five clouds. A systematic decrease of  polarisation efficiency is seen with increasing $A_{V}$ in the clouds. Only a small subset of the interstellar dust grains causes the polarisation of the background starlight \citep{1995ApJ...448..748G}. If the dominant population of the asymmetric dust grains lying along the line of sight are misaligned with the local magnetic field, the polarisation efficiency could decrease with the extinction. The drop in the polarisation efficiency can also happen if the dust grains in the cold dark clouds are larger and more spherical than those in the outer regions \citep{1993AJ....105.1010V}.

In the lower panel of Fig. \ref{Fig:poleff}, we show the variation of P with $A_{V}$. The solid line represents the observational upper limit of P/$A_{V}$ = 3\% $mag^{-1}$. This plot suggests that the dust grains sampled by this study lie below the line representing optimum alignment efficiency. This indicates that the dust content in these objects should be similar to the diffuse interstellar medium. The relationship between P and $A_{V}$ provides information on the polarising capability of the grains along the line of sight and/or about the orientation of the magnetic field \citep{1995ApJ...448..748G}. In all of the clouds except L1521F, the P values lie below the observational upper limit of the amount of the polarisation possible for a particular extinction. The presence of the low grain alignment efficiency or tangled magnetic field could reduce the P values for a given extinction. In this plot, L1521F seems to approach the border of the optimum alignment between dust grains and the magnetic field lines (within the error bars). The mean value of P (= 4.5\%) in L1521F is the highest among the clouds studied here. A relatively small value of the line-of-sight magnetic field strength \citep{2008ApJ...680..457T} in L1521F indicates that the space orientation of the magnetic field is almost in the plane of the sky. Hence the dust grains aligned to the magnetic field show a high polarisation efficiency. The stronger plane-of-the-sky magnetic field strength indicates the reduced projection effect in this cloud. The large offset of $\sim 60\degree$ between the minor axis of L1521F core \citep[inferred from the 1.2 mm MAMBO mapping;][]{2008A&A...487..993K} and the envelope magnetic field cannot be explained based on the magnetic models of cloud evolution. However, the $13\degree$ offset between the minor axis inferred from the optically selected core \citep{1999ApJS..123..233L} and the outer magnetic field suggests that at least the initial evolution of the cloud might have been mediated by the magnetic field. The denser inner part of the core seen in MAMBO mapping might have got decoupled from the outer part as the core evolved. Mapping of the core magnetic field in submillimetre wavelengths would give more insight into the evolution history of the cloud.

\section{Conclusions}\label{Conclusions}
 
We present the results of our optical polarisation measurements of stars projected on five dense cores namely, IRAM 04191, L1521F, L328, L673-7, and L1014 which are found to harbor VeLLOs detected by the \textit{Spitzer Space Telescope}. The polarimetric measurements of stars projected on IRAM 04191, L328, L673-7, and L1014 were made in R-band and measurements of stars towards L1521F were made in V-band. The results obtained are summarised below.

\begin{itemize}
\item The angular offsets between the envelope magnetic field direction (inferred from optical polarisation measurements) and the outflow position angles from the VeLLOs for IRAM 04191, L1521F, L328, L673-7, and L1014 are found to be 84$\degree$, 53$\degree$, 24$\degree$, 8$\degree$, and 15$\degree$, respectively. The mean value of the offsets for the five clouds is $\sim37\degree$. If we exclude IRAM 04191 which shows an almost perpendicular orientation, the mean value becomes $\sim25\degree$. Magnetic field maps of a higher number of cores with VeLLO sources are required to make statistically significant conclusions.

\item The angular offset between the envelope magnetic field direction and the minor axis (inferred based on the 850$\mu$m and 1.2~mm dust continuum emission maps available in the literature) of IRAM 04191, L1521F, L673-7, and L1014 are 82$\degree$, 60$\degree$, 47$\degree$, and 55$\degree$, respectively. The mean value of the offsets for these cores is found to be $\sim60\degree$. The values of the angular offset between $\theta^{opt}_{B}$ and $\theta^{opt}_{min}$ for L1521F, L328, and L1014 are 13$\degree$, 1$\degree$, and 17$\degree$, respectively. The mean value of the offsets for these cores is found to be $\sim10\degree$. Thus, the outer magnetic fields are found to be more parallel to the minor axes inferred from the optically fitted ellipse to the cores than the minor axes inferred from the submillimetre continuum emission maps.

\item The angular offsets between the envelope and the inner magnetic field (inferred from the submillimetre data obtained using SCUPOL) in IRAM 04191 is found to be 68$\degree$. However, the inner magnetic field is found to be nearly aligned with the projected position angles of the minor axis, the rotation axis of the cloud, and the outflow from the IRAM 04191-IRS.

\item The plane-of-the-sky component of the magnetic field strength estimated for IRAM 04191, L1521F, L328, and L1014 are  $\sim28\mu$G, $\sim24\mu$G, $\sim15\mu$G, and $\sim20\mu$G, respectively. The typical uncertainty ($\sigma B_{POS}$) in the magnetic field strength is estimated to be $\sim0.5B_{POS}$. At the cloud scale, the dependence of alignment between the magnetic field and the outflow direction on the magnetic field strengths is not very apparent in the clouds studied here.

\item The dust grains sampled in this study are found to be located below the line representing optimum alignment efficiency. This suggest that the material composing these objects are similar to the diffused ISM. However, L1521F reaches the border of optimum alignment between dust grains and the magnetic field lines (within the error bars). In L1521F, the high degree of polarisation may be due to the higher efficiency of the grains aligned with the magnetic field.

\end{itemize}

Magnetic field maps of a higher number of cores with VeLLOs are required to arrive at statistically significant conclusions. It would also be interesting to probe the core scale magnetic field directions of clouds with VeLLOs using submillimetre polarimetry. This would allow us to study the correlation between outflows and core scale magnetic fields and the relative orientation between the envelope and inner magnetic fields. The results would provide great insight into the formation and evolution of very low-mass and substellar-mass sources.

\begin{table}  %{f}
\begin{minipage}{80mm}
\caption{Optical polarisation results of the stars observed in the direction of five dense cores with VeLLOs.}\label{tab:pol_res}
\begin{tabular}{llllr}\hline
Star  & $\alpha$ (J2000)  & $\delta$ (J2000)  & P $\pm$ $\sigma_P$ & $\theta$ $\pm$ $\sigma_{\theta}$  \\ 
 Id  &($\degree$)&($\degree$)& (\%) &($\degree$) \\\hline  
\hline
\multicolumn{5}{c}{\bf{IRAM 04191}}\\
1 &65.306771&	$+$15.382839&2.1$\pm$0.6&	101$\pm$8 \\
2 &65.321500&	$+$15.383724&1.3$\pm$0.2&	104$\pm$4 \\
3 &65.322121&	$+$15.620441&1.3$\pm$0.2&	117$\pm$5 \\
4 &65.342513&	$+$15.639670&2.4$\pm$0.8&	112$\pm$10\\
5 &65.343609&	$+$15.411406&2.1$\pm$0.7&	104$\pm$10\\
6 &65.349026&	$+$15.629446&2.4$\pm$0.3&	106$\pm$5 \\
7 &65.361083&	$+$15.418915&1.8$\pm$0.7&	99 $\pm$11\\
8 &65.395908&	$+$15.593961&2.1$\pm$0.3&	121$\pm$4 \\
9 &65.406631&	$+$15.583215&5.5$\pm$0.7&	125$\pm$4 \\
10&65.408752&	$+$15.654627&1.4$\pm$0.1&	119$\pm$1 \\
\multicolumn{5}{c}{}\\
11&65.537750&	$+$15.439396&2.3$\pm$0.1&	97 $\pm$1 \\
12&65.547373&	$+$15.417146&2.2$\pm$0.2&	90 $\pm$3 \\
13&65.551323&	$+$15.581253&1.2$\pm$0.1&	123$\pm$3 \\
$14^{a}$&65.557352&	$+$15.424945&2.2$\pm$0.1&	90 $\pm$1 \\
15&65.564020&	$+$15.594937&2.1$\pm$0.7&	110$\pm$9 \\
16&65.566471&	$+$15.441818&4.6$\pm$1.1&	133$\pm$7 \\
17&65.600283&	$+$15.401129&3.1$\pm$0.4&	125$\pm$4 \\	 
18&65.634927&	$+$15.584133&1.1$\pm$0.2&	125$\pm$7 \\
19&65.641876&	$+$15.561810&2.4$\pm$0.8&	145$\pm$9 \\
20&65.713315&	$+$15.465842&1.3$\pm$0.5&	119$\pm$11\\
\multicolumn{5}{c}{}\\
21&65.720609&	$+$15.470545&1.9$\pm$0.2&	113$\pm$4 \\
22&65.724916&	$+$15.473766&1.4$\pm$0.2&	108$\pm$4 \\
23&65.728694&	$+$15.531259&1.7$\pm$0.4&	107$\pm$8 \\
24&65.743244&	$+$15.490355&1.4$\pm$0.4&	114$\pm$9 \\
25&65.759093&	$+$15.476655&3.5$\pm$0.9&	118$\pm$8 \\\hline\hline
\multicolumn{5}{c}{\bf{L1521F (data is taken using V-band)}}\\
1 &66.992217&	$+$26.745798&	4.1$\pm$0.9&	78 $\pm$	7 \\
2 &67.005072&	$+$26.720570&	2.6$\pm$0.6&	41 $\pm$	7 \\
3 &67.019487&	$+$26.791056&	3.1$\pm$1.2&	67 $\pm$	12 \\
4 &67.020278&	$+$26.835775&	6.3$\pm$1.4&	28 $\pm$	6 \\
5 &67.020330&	$+$26.690828&	1.9$\pm$0.4&	29 $\pm$	6 \\
6 &67.022458&	$+$26.695829&	2.2$\pm$0.4&	32 $\pm$	6 \\
7 &67.028105&	$+$26.776918&	3.8$\pm$0.8&	54 $\pm$	6 \\
8 &67.034057&	$+$26.807844&	6.1$\pm$1.2&	13 $\pm$	6 \\
9 &67.039305&	$+$26.719938&	3.5$\pm$0.6&	40 $\pm$	5 \\
10&67.048615&	$+$26.736067&	1.5$\pm$0.4&	42 $\pm$	7 \\
 \multicolumn{5}{c}{}\\
11&67.073492&	$+$26.836891&	3.7$\pm$1.8&	55 $\pm$	14\\
12&67.079859&	$+$26.796986&	1.9$\pm$0.5&	34 $\pm$	8 \\
13&67.082504&	$+$26.634434&	6.5$\pm$1.1&	31 $\pm$	4 \\
14&67.111749&	$+$26.992775&	2.4$\pm$0.1&	18 $\pm$	2 \\
15&67.113738&	$+$26.993118&	2.7$\pm$0.2&	18 $\pm$	2 \\
16&67.136871&	$+$26.938223&	6.1$\pm$2.3&	45 $\pm$	12\\
17&67.158933&	$+$26.613392&	4.6$\pm$0.5&	23 $\pm$	3 \\
18&67.170430&	$+$26.928179&	2.9$\pm$0.2&	12 $\pm$	2 \\
19&67.179662&	$+$26.961123&	3.5$\pm$0.3&	8  $\pm$	2 \\
20&67.199998&	$+$26.956436&	6.3$\pm$2.2&	173$\pm$	10\\
 \multicolumn{5}{c}{}\\
%\end{tabular}
%\end{minipage}
%\end{table}
%\begin{table}
%\centering
%\begin{minipage}{80mm}
%\begin{tabular}{llllr}\hline
%Star  & $\alpha$ (J2000)  & $\delta$ (J2000)  & P $\pm$ $\sigma_P$ & $\theta$ $\pm$ $\sigma_{\theta}$  \\\hline
21&67.208424&	$+$27.037630&	3.1$\pm$1.4&	37 $\pm$	12\\
22&67.221287&	$+$27.025188&	2.7$\pm$0.4&	21 $\pm$	5 \\
23&67.227841&	$+$26.638966&	2.2$\pm$0.2&	26 $\pm$	3 \\
24&67.234556&	$+$26.696199&	7.1$\pm$0.6&	23 $\pm$	2 \\
25&67.261526&	$+$26.628933&	2.1$\pm$0.6&	1  $\pm$	8 \\
26&67.266799&	$+$26.683596&	7.7$\pm$0.9&	26 $\pm$	3 \\
27&67.275536&	$+$26.830238&	4.7$\pm$1.3&	16 $\pm$	8 \\
28&67.292543&	$+$26.873404&	4.3$\pm$0.8&	16 $\pm$	5 \\
29&67.293557&	$+$26.862020&	2.1$\pm$0.7&	17 $\pm$	8 \\
30&67.299197&	$+$26.624643&	4.9$\pm$0.7&	13 $\pm$	4 \\
 \multicolumn{5}{c}{}\\
31&67.309591&	$+$26.888031&	1.7$\pm$0.4&	14 $\pm$	7 \\
32&67.318660&	$+$26.809628&	2.3$\pm$0.9&	21 $\pm$	12\\
33&67.319464&	$+$26.821066&	4.1$\pm$1.7&	31 $\pm$	12\\
34&67.321832&	$+$26.667566&	2.2$\pm$0.3&	26 $\pm$	4 \\
35&67.326550&	$+$26.889544&	7.9$\pm$2.3&	8  $\pm$	8 \\
36&67.338066&	$+$26.840038&	2.2$\pm$0.4&	5  $\pm$	4 \\
37&67.344587&	$+$26.641081&	3.2$\pm$0.4&	18 $\pm$	3 \\
38&67.373060&	$+$26.881413&	3.4$\pm$0.3&	16 $\pm$	2 \\
39&67.387662&	$+$26.878212&	2.8$\pm$0.6&	11 $\pm$	6 \\
40&67.404212&	$+$26.845400&	1.9$\pm$0.8&	19 $\pm$	12\\\hline\hline
\end{tabular}
\end{minipage}
\end{table}
\begin{table}
\centering
\begin{minipage}{80mm}
%\contcaption{}
\begin{tabular}{llllr}\hline
Star  & $\alpha$ (J2000)  & $\delta$ (J2000)  & P $\pm$ $\sigma_P$ & $\theta$ $\pm$ $\sigma_{\theta}$  \\\hline
\multicolumn{5}{c}{\bf{L328 (data is taken using $R_{kc}$-band)}} \\
1&   274.167511&  -18.033754&  0.7$\pm$  0.3&  152 $\pm$  13 \\
2&   274.171753&  -18.057564&  2.1$\pm$  0.3&  106 $\pm$  4  \\
3&   274.172913&  -18.060534&  0.9$\pm$  0.2&  109 $\pm$  8  \\
4&   274.173615&  -18.039865&  3.7$\pm$  0.3&  100 $\pm$  2  \\
5&   274.178345&  -18.052841&  4.2$\pm$  0.5&  138 $\pm$  3  \\
6&   274.183197&  -18.072594&  0.7$\pm$  0.3&  165 $\pm$  12 \\
7&   274.185089&  -18.030083&  0.8$\pm$  0.1&  16  $\pm$  6  \\
8&   274.185760&  -18.031004&  0.8$\pm$  0.1&  6   $\pm$  5  \\
9&   274.189575&  -17.997437&  3.3$\pm$  0.4&  19  $\pm$  4  \\
10&  274.196747&  -17.996161&  1.2$\pm$  0.2&  179 $\pm$  5  \\
\multicolumn{5}{c}{}\\	
11&  274.201721&  -18.041695&  1.1$\pm$  0.2&  11  $\pm$  5 \\
12&  274.202240&  -18.002193&  2.4$\pm$  0.1&  6   $\pm$  1 \\
13&  274.202362&  -18.010677&  1.4$\pm$  0.1&  5   $\pm$  2 \\
14&  274.206909&  -18.051521&  0.6$\pm$  0.1&  143 $\pm$  9 \\
15&  274.209747&  -18.012936&  1.4$\pm$  0.2&  19  $\pm$  4 \\
16&  274.210693&  -18.040684&  0.8$\pm$  0.3&  108 $\pm$  12\\
17&  274.211304&  -18.035072&  0.9$\pm$  0.3&  88  $\pm$  11\\
18&  274.214386&  -18.106380&  3.9$\pm$  0.3&  7   $\pm$  2 \\
19&  274.214935&  -18.036556&  1.1$\pm$  0.3&  5   $\pm$  8 \\
20&  274.215698&  -17.991894&  1.4$\pm$  0.3&  11  $\pm$  6 \\
\multicolumn{5}{c}{}\\	
21&  274.218536&  -18.021330&  1.2$\pm$  0.4&  38  $\pm$  10\\
22&  274.224579&  -18.112123&  2.6$\pm$  0.2&  174 $\pm$  3 \\
23&  274.227478&  -18.021618&  1.5$\pm$  0.5&  150 $\pm$  10\\
24&  274.230591&  -18.098825&  1.8$\pm$  0.6&  144 $\pm$  10\\
25&  274.230621&  -18.098833&  1.2$\pm$  0.4&  116 $\pm$  10\\
26&  274.230713&  -18.025410&  1.9$\pm$  0.4&  90  $\pm$  6 \\
27&  274.233368&  -18.022760&  0.9$\pm$  0.3&  170 $\pm$  9 \\
28&  274.234161&  -18.102222&  1.1$\pm$  0.2&  106 $\pm$  6 \\
29&  274.234253&  -18.102201&  2.7$\pm$  0.3&  153 $\pm$  4 \\
30&  274.234955&  -18.010530&  0.7$\pm$  0.2&  94  $\pm$  12\\
\multicolumn{5}{c}{}\\	
$31^{b}$&  274.235931&  -18.022663&  1.2$\pm$  0.2&  162 $\pm$  5 \\
32&  274.238403&  -18.006271&  0.9$\pm$  0.3&  140 $\pm$  11\\
33&  274.239685&  -18.094095&  0.7$\pm$  0.2&  97  $\pm$  8 \\
34&  274.239807&  -18.094038&  1.5$\pm$  0.3&  166 $\pm$  5 \\
35&  274.240082&  -18.013145&  1.6$\pm$  0.2&  162 $\pm$  4 \\
36&  274.247070&  -17.987055&  2.9$\pm$  0.6&  166 $\pm$  5 \\
37&  274.247498&  -18.119047&  2.3$\pm$  0.4&  116 $\pm$  5 \\
38&  274.252808&  -18.083229&  0.6$\pm$  0.2&  179 $\pm$  7 \\
39&  274.253815&  -18.091532&  1.1$\pm$  0.2&  110 $\pm$  5 \\
40&  274.253845&  -18.091520&  0.8$\pm$  0.4&  165 $\pm$  13\\
\multicolumn{5}{c}{}\\	
41&  274.256653&  -18.087948&  0.9$\pm$  0.3&  1   $\pm$  11\\
42&  274.257904&  -18.095560&  1.2$\pm$  0.4&  53  $\pm$  9 \\
43&  274.258087&  -18.046804&  2.1$\pm$  0.3&  168 $\pm$  5 \\
44&  274.259033&  -18.006433&  0.6$\pm$  0.2&  40  $\pm$  9 \\
45&  274.259094&  -18.006470&  1.3$\pm$  0.1&  65  $\pm$  3 \\
46&  274.260193&  -18.091240&  0.8$\pm$  0.1&  177 $\pm$  4 \\
47&  274.260193&  -18.091282&  1.1$\pm$  0.2&  14  $\pm$  6 \\
48&  274.260773&  -18.061516&  1.5$\pm$  0.4&  19  $\pm$  7 \\
49&  274.267365&  -18.005779&  1.4$\pm$  0.2&  93  $\pm$  6 \\
50&  274.267426&  -18.005766&  1.1$\pm$  0.2&  104 $\pm$  8 \\
\multicolumn{5}{c}{}\\	
%\end{tabular}
%\end{minipage}
%\end{table}
%\begin{table}
%\centering
%\begin{minipage}{80mm}
%\contcaption{}
%\begin{tabular}{llllr}\hline
%Star  & $\alpha$ (J2000)  & $\delta$ (J2000)  & P $\pm$ $\sigma_P$ & $\theta$ $\pm$ $\sigma_{\theta}$  \\\hline
%\multicolumn{5}{c}{}\\	
51&  274.267639&  -18.101971&  1.1$\pm$  0.4&  13  $\pm$  12\\
52&  274.270416&  -18.020172&  1.9$\pm$  0.4&  37  $\pm$  6 \\
53&  274.274933&  -18.103592&  1.1$\pm$  0.3&  142 $\pm$  10\\
54&  274.275513&  -18.061523&  0.3$\pm$  0.1&  175 $\pm$  12\\
55&  274.275604&  -18.076469&  0.9$\pm$  0.3&  25  $\pm$  10\\
56&  274.276947&  -17.971384&  0.8$\pm$  0.4&  73  $\pm$  13\\
57&  274.278168&  -18.035854&  0.5$\pm$  0.2&  87  $\pm$  9 \\
58&  274.280487&  -18.098484&  0.8$\pm$  0.2&  171 $\pm$  7 \\
59&  274.281616&  -18.011997&  0.7$\pm$  0.3&  88  $\pm$  12\\
60&  274.284576&  -18.041805&  0.4$\pm$  0.1&  25  $\pm$  13\\
\multicolumn{5}{c}{}\\	
61&  274.284576&  -18.041857&  1.1$\pm$  0.2&  45  $\pm$  5 \\
62&  274.285583&  -18.026943&  2.5$\pm$  0.4&  55  $\pm$  4 \\
63&  274.286591&  -18.001257&  1.7$\pm$  0.3&  66  $\pm$  5 \\
64&  274.288147&  -18.076719&  1.3$\pm$  0.3&  140 $\pm$  6 \\
\end{tabular}
\end{minipage}
\end{table}
\begin{table}
\centering
\begin{minipage}{80mm}
%\contcaption{}
\begin{tabular}{llllr}\hline
Star  & $\alpha$ (J2000)  & $\delta$ (J2000)  & P $\pm$ $\sigma_P$ & $\theta$ $\pm$ $\sigma_{\theta}$  \\\hline
%\multicolumn{5}{c}{}\\	
65&  274.288208&  -18.054703&  0.9$\pm$  0.3&  40  $\pm$  10\\
66&  274.288208&  -18.054787&  1.1$\pm$  0.4&  6   $\pm$  10\\
67&  274.298462&  -17.998753&  2.7$\pm$  0.3&  61  $\pm$  3 \\
68&  274.301117&  -18.008818&  2.8$\pm$  0.1&  65  $\pm$  1 \\
69&  274.301208&  -18.067600&  1.1$\pm$  0.1&  30  $\pm$  5 \\
70&  274.301208&  -18.010597&  1.2$\pm$  0.2&  62  $\pm$  4 \\
\multicolumn{5}{c}{}\\	
71&  274.306458&  -18.067343&  0.5$\pm$  0.1&  102 $\pm$  5 \\
72&  274.310760&  -18.036221&  1.4$\pm$  0.3&  85  $\pm$  6 \\
73&  274.311493&  -18.034235&  1.2$\pm$  0.2&  75  $\pm$  6 \\
74&  274.314301&  -18.030848&  1.1$\pm$  0.2&  60  $\pm$  5 \\
75&  274.315125&  -18.034849&  0.8$\pm$  0.3&  153 $\pm$  9 \\
76&  274.315552&  -18.040340&  0.9$\pm$  0.2&  87  $\pm$  7 \\
77&  274.323273&  -18.044617&  0.9$\pm$  0.2&  46  $\pm$  6 \\ \hline\hline
\multicolumn{5}{c}{\bf{L673-7}}\\
1 &	290.299498&$+$11.354861&	1.4$\pm$0.1&	64 $\pm$3\\	
2 &	290.306563&$+$11.292442&	1.4$\pm$0.3&	111$\pm$6 \\	
3 &	290.309463&$+$11.271273&	1.7$\pm$0.7&	113$\pm$12\\	
4 &	290.314065&$+$11.253704&	0.8$\pm$0.2&	66 $\pm$7\\	
5 &	290.320294&$+$11.379147&	3.7$\pm$0.8&	48 $\pm$6\\	
6 &	290.322640&$+$11.269587&	0.6$\pm$0.2&	133$\pm$10\\	
7 &	290.337911&$+$11.377842&	0.5$\pm$0.1&	165$\pm$6\\	
8 &	290.338487&$+$11.332059&	1.3$\pm$0.2&	135$\pm$5\\	
9 &	290.340009&$+$11.386873&	1.5$\pm$0.6&	50 $\pm$10\\	
10&	290.356531&$+$11.362033&	2.3$\pm$0.7&	21 $\pm$8\\
\multicolumn{5}{c}{}\\	
11&	290.370054&$+$11.379834&	1.6$\pm$0.2&	156$\pm$5\\	
12&	290.374782&$+$11.329640&	0.8$\pm$0.4&	56 $\pm$12\\	
13&	290.376037&$+$11.348548&	0.6$\pm$0.2&	61 $\pm$10\\	
14&	290.377778&$+$11.453568&	2.6$\pm$1.1&	88 $\pm$10\\	
15&	290.384464&$+$11.232626&	2.4$\pm$0.4&	104$\pm$6\\	
16&	290.387623&$+$11.284531&	1.9$\pm$0.2&	141$\pm$4\\	
17&	290.394228&$+$11.331353&	4.2$\pm$0.7&	21 $\pm$4\\	
18&	290.398035&$+$11.255313&	0.6$\pm$0.2&	136$\pm$10\\	
19&	290.401768&$+$11.282085&	0.9$\pm$0.1&	124$\pm$4\\	
20&	290.405570&$+$11.448447&	1.7$\pm$0.4&	57 $\pm$6\\
\multicolumn{5}{c}{}\\	
21&	290.425520&$+$11.306265&	2.4$\pm$0.3&	95 $\pm$4\\	
22&	290.440832&$+$11.473078&	2.4$\pm$0.6&	34 $\pm$8\\	
23&	290.455609&$+$11.282369&	0.7$\pm$0.2&	5  $\pm$8\\	
24&	290.461728&$+$11.334425&	3.9$\pm$0.3&	1  $\pm$2\\	
25&	290.469662&$+$11.423946&	2.6$\pm$0.3&	60 $\pm$3\\
26&	290.471921&$+$11.373240&	2.6$\pm$0.4&	54 $\pm$4\\	
27&	290.473356&$+$11.419109&	3.1$\pm$0.8&	79 $\pm$7\\	
28&	290.478498&$+$11.391085&	2.6$\pm$0.5&	66 $\pm$5\\	
29&	290.478786&$+$11.350402&	0.8$\pm$0.3&	44 $\pm$11\\	
30&	290.482362&$+$11.304210&	2.8$\pm$0.3&	150$\pm$3\\	
\multicolumn{5}{c}{}\\
31&	290.482767&$+$11.265624&	1.2$\pm$0.2&	153$\pm$4\\	
32&	290.508964&$+$11.271605&	0.6$\pm$0.1&	29 $\pm$5\\	
%\end{tabular}
%\end{minipage}
%\end{table}
%\begin{table}
%\centering
%\begin{minipage}{80mm}
%\contcaption{}
%\begin{tabular}{llllr}\hline
%Star  & $\alpha$ (J2000)  & $\delta$ (J2000)  & P $\pm$ $\sigma_P$ & $\theta$ $\pm$ $\sigma_{\theta}$  \\\hline	
33&	290.509317&$+$11.291866&	1.3$\pm$0.5&	61 $\pm$10\\	
34&	290.517582&$+$11.293197&	2.3$\pm$0.3&	3  $\pm$4\\	
35&	290.527966&$+$11.334507&	2.1$\pm$0.4&	42 $\pm$6\\	
36&	290.530922&$+$11.286077&	3.3$\pm$0.6&	139$\pm$5\\	
37&	290.533773&$+$11.336944&	2.5$\pm$0.4&	69 $\pm$5\\\hline\hline
\end{tabular}
\end{minipage}
\end{table}
\begin{table}
\centering
\begin{minipage}{80mm}
%\contcaption{}
\begin{tabular}{llllr}\hline
Star  & $\alpha$ (J2000)  & $\delta$ (J2000)  & P $\pm$ $\sigma_P$ & $\theta$ $\pm$ $\sigma_{\theta}$  \\\hline	
\multicolumn{5}{c}{\bf{L1014}}\\
  1&  320.905731& $+$49.952625&     1.2$\pm$    0.1& 15$\pm$   2\\
  2&  320.905731& $+$49.971397&     0.6$\pm$    0.1& 17$\pm$   5\\
  3&  320.908173& $+$49.940659&     1.1$\pm$    0.2& 76$\pm$   5 \\
  4&  320.909210& $+$49.992123&     1.4$\pm$    0.3& 26$\pm$   7\\
  5&  320.920593& $+$49.958851&     2.7$\pm$    1.1& 14$\pm$  12\\
  $6^{c}$&  320.936371& $+$49.996620&     2.1$\pm$    0.2& 19$\pm$   2\\
  7&  320.941925& $+$49.967125&     2.2$\pm$    0.3& 81$\pm$   4\\
  8&  320.946442& $+$49.912392&     0.5$\pm$    0.2& 25$\pm$  11\\
  9&  320.946747& $+$50.022636&     1.2$\pm$    0.1& 31$\pm$   4\\
  10&  320.951630& $+$49.929806&     2.1$\pm$    0.3&  2$\pm$   4\\
  \multicolumn{5}{c}{}\\
  11&  320.951935& $+$49.926811&     3.4$\pm$    0.4&  9$\pm$   4\\
  12&  320.954742& $+$49.930172&     2.7$\pm$    0.6& 23$\pm$   6\\
  13&  320.957092& $+$49.934036&     0.8$\pm$    0.1& 26$\pm$   5\\
  $14^{d}$&  320.958008& $+$50.030022&     0.4$\pm$    0.1& 66$\pm$   6\\
  15&  320.964233& $+$49.932388&     1.3$\pm$    0.3&  6$\pm$   7\\
  16&  320.970154& $+$49.916588&     0.7$\pm$    0.3& 14$\pm$  12\\
  17&  320.973236& $+$49.922066&     1.7$\pm$    0.5& 12$\pm$   8\\
  18&  320.976227& $+$49.932659&     1.1$\pm$    0.1& 18$\pm$   4\\
  19&  320.979645& $+$50.020786&     1.8$\pm$    0.5& 12$\pm$   8\\
  20&  320.981354& $+$49.921574&     0.8$\pm$    0.1& 19$\pm$   6\\
  \multicolumn{5}{c}{}\\
  21&  320.985291& $+$49.942341&     1.6$\pm$    0.3& 97$\pm$   7\\
  22&  320.987671& $+$49.928112&     1.4$\pm$    0.2&  3$\pm$   3\\
  23&  320.989716& $+$49.929646&     1.1$\pm$    0.2& 17$\pm$   6\\
  24&  320.993195& $+$50.017593&     2.1$\pm$    0.2& 22$\pm$   4\\
  $25^{e}$&  320.993805& $+$49.969070&     1.2$\pm$    0.2& 55$\pm$   6\\
  26&  321.001984& $+$50.031876&     1.1$\pm$    0.3&  8$\pm$  10\\
  27&  321.003296& $+$49.928627&     1.3$\pm$    0.3& 11$\pm$   8\\
  28&  321.020874& $+$50.046665&     1.1$\pm$    0.2& 30$\pm$   6\\
  29&  321.024567& $+$49.971264&     1.2$\pm$    0.5& 22$\pm$  11\\
  30&  321.036407& $+$49.933159&     0.9$\pm$    0.3& 40$\pm$   9\\
  \multicolumn{5}{c}{}\\
  31&  321.040283& $+$50.010227&     1.5$\pm$    0.5& 12$\pm$  10\\
  32&  321.042633& $+$50.025105&     3.2$\pm$    1.4& 29$\pm$  13\\
  33&  321.062622& $+$49.973835&     1.1$\pm$    0.3&  4$\pm$   8\\
  34&  321.063293& $+$50.003963&     2.8$\pm$    0.6& 17$\pm$   6\\
  35&  321.068481& $+$50.010017&     1.6$\pm$    0.6& 49$\pm$  10\\
  36&  321.075165& $+$50.008911&     1.1$\pm$    0.5& 35$\pm$  12\\
  37&  321.079315& $+$49.951550&     0.7$\pm$    0.3&  7$\pm$   9\\
  38&  321.084381& $+$50.032486&     1.2$\pm$    0.2& 22$\pm$   5\\
  \hline
\hline
\end{tabular}\\

a) IRAS 04193$+$1518 (YSO) \\
b) [M81] I--651 (Emission-line Star) \\
c) K2.5 2MASS J21234472$+$4959478 (Emission-line Star) \\
d) TYC 3598-2027--1\\
e) K6.5 2MASS J21235851$+$4958086 (Emission-line Star)\\ 
\end{minipage} \\
\end{table}   %  pol results

%=====================================================
%******************************************************************************************
%***********************ACKNOWLEDGEMENT****************************************************
%******************************************************************************************
\section{Acknowledgements}
The authors are very grateful to the anonymous referee for the constructive comments and suggestions, which helped considerably to improve the content of the manuscript. This research has made use of the SIMBAD database, operated at CDS, Strasbourg, France. We also acknowledge the use of NASA's \textit{SkyView} facility (http://skyview.gsfc.nasa.gov) located at NASA Goddard Space Flight Center. CWL was supported by Basic Science Research Program though the National Research Foundation of Korea (NRF) funded by the Ministry of Education, Science, and Technology (NRF-2013R1A1A2A10005125) and also by the global research collaboration of Korea Research Council of Fundamental Science \& Technology (KRCF). MT is supported by JSPS fund (No. 22000005). SD is supported by a Marie-Curie Intra European Fellowship under the European Community's Seventh Framework Program FP7/2007-2013 grant agreement no 627008.

%======================================================================================
\bibliographystyle{aa}
\bibliography{VeLLO_ref}
\clearpage
%======================================================================================
%\input{table4}   %  pol results

%\input{table6}     % Info of stars

\begin{table*}
\centering
\begin{minipage}{\textwidth}
\caption{Mean magnetic field position angles and angular offsets between magnetic field, cloud minor axis, and outflows.}\label{tab:offsets}
\begin{tabular}{llccccccccc}
\hline
Cloud Id.

& $\theta^{opt}_{B}$
& $\theta^{sub}_{B}$		
& $|\theta^{opt}_{B}-\theta^{sub}_{B}|$
& $|\theta^{opt}_{B}-\theta_{min}|$
& $|\theta^{sub}_{B}-\theta_{min}|$
& $|\theta^{opt}_{B}-\theta_{out}|$
& $|\theta^{sub}_{B}-\theta_{out}|$
& $|\theta_{out}-\theta_{min}|$
& $\sigma_{\theta^{opt}_{B}}$
\\

&(deg)	
&($^{\circ}$)	
&($^{\circ}$)
&($^{\circ}$)	
&($^{\circ}$)	
&($^{\circ}$)	
&($^{\circ}$)
&($^{\circ}$)\\

(1)&(2)&(3)&(4)&(5)&(6)&(7)&(8)&(9)&(10)\\
\hline
 IRAM 04191 & 112 	&44	 	&68		&82  	& 14 	& 84 			& 16  		&2 &  13   \\
 L1521F     & 22  	&--	 	&--		&60		& -- 	& 53 			& -- 		&07 & 13  \\
 L328       & 44  	&--	 	&--		&--		& -- 	& 24 			& -- 		&-- & 21  \\
 L673-7     & 47  	&--	 	&--		&47		& -- 	& 8  			& -- 		&55 & 48 \\
 L1014      & 15  	&--	 	&--		&55		& -- 	& 15 			& --  		&40 & 13  \\
\hline
\end{tabular}

$opt:$ Optical data; $sub:$ Submm data; $min:$ Minor axis of the clouds; $out:$ Outflow direction from the VeLLO
\end{minipage}
\end{table*}

%======================================================================================

\label{lastpage}
\end{document}